\documentclass[manuscript]{aastex62}
\usepackage{graphicx}
\usepackage{csvsimple}
\usepackage{float}
\usepackage{placeins}
\usepackage{bm}
\accepted{by ApJS February 15, 2019}

\shorttitle{}
\shortauthors{}

\begin{document}
\title{The Period-Luminosity Relations of Red Supergiants in M33 and M31} 

\correspondingauthor{Bi-Wei Jiang}
\email{bjiang@bnu.edu.cn}

\author{Yi Ren}
\affiliation{Department of Astronomy, Beijing Normal University, Beijing 100875, China}

\author{Bi-Wei Jiang}
\affiliation{Department of Astronomy, Beijing Normal University, Beijing 100875, China}

\author{Ming Yang}
\affiliation{Institute for Astronomy, Astrophysics, Space Applications \& Remote Sensing, National Observatory of Athens, Vas. Pavlou and I. Metaxa, Penteli 15236, Greece}

\author{Jian Gao}
\affiliation{Department of Astronomy, Beijing Normal University, Beijing 100875, China}


\begin{abstract}
Based on previously selected preliminary samples of Red Supergiants (RSGs) in M33 and M31, the foreground stars and luminous Asymptotic Giant Branch stars (AGBs) are further excluded, which leads to the samples of 717 RSGs in M33 and 420 RSGs in M31. With the time-series data from the iPTF survey spanning nearly 2000 days, the period and amplitude of RSGs are analyzed. According to the lightcurves characteristics, they are classified into four categories in which 84 and 56 objects in M33 and M31 respectively are semi-regular variables. For these semi-regular variables, the pulsation mode is identified by comparing with the theoretical model, which yielded 19 (7) sources in the first overtone mode in M33 (M31), and the other 65 (49) RSGs in M33 (M31) in the fundamental mode. The period-luminosity (P-L) relation is analyzed for the RSGs in the fundamental mode. It is found the P-L relation is tight in the infrared, i.e. the 2MASS $JHK_{\rm S}$ bands and the short-wavelength bands of Spitzer. Meanwhile, the inhomogeneous extinction causes the P-L relation scattering in the $V$ band, and the dust emission causes the less tight P-L relation in the Spitzer/[8.0] and [24] bands. The derived P-L relations in the 2MASS/$K_{\rm S}$ band are in agreement with those of RSGs in SMC, LMC and the Milky Way within the uncertainty range. It is found that the number ratio of RSGs pulsating in the fundamental mode to the first overtone mode increases with metallicity.

\end{abstract}

\keywords{stars: late-type stars: oscillations stars: variables: other supergiants}


\section{Introduction} \label{sec:intro}

Red Supergiants (RSGs) are Population \uppercase\expandafter{\romannumeral1} stars in helium burning stage with a mass range of about 9-27${\rm M_\odot}$. They have relatively low effective temperature of $\sim 3000-4000$\,K and the corresponding spectral types of late K- to M-types \citep{2008IAUS..250...97M}. Their radius is large, with the maximum radius being $\sim 1500 {R_\odot}$ \citep{2005ApJ...628..973L}, so that they have low surface gravity of $  \log \rm g< 1.0$, but high luminosity of $\sim 3\,500-630\,000 {L_\odot}$ \citep{2008IAUS..250...97M, 2016ApJ...826..224M}.

RSG is a phase of very significant mass loss \citep{1996ApJ...468..842S}. The Reimers law \citep{1975MSRSL...8..369R, 1978A&A....70..227K} is an empirical relation between mass loss rate (MLR), stellar luminosity, radius and mass derived from a sample of red giants and RSGs: $\dot{M} = 5.5 * 10^{-13} {L_\odot R_\odot /M_\odot}$. \citet{2005A&A...438..273V} also derived the relationship between MLR and stellar parameters, but based on the samples of RSGs and oxygen-rich Asymptotic Giant Branch stars (AGBs) in Large Magellanic Cloud (LMC): $\log (\dot{M}) = -5.5 + 1.05 \log (L/10000{L_\odot}) - 6.3 \log (T_{\rm eff}/3500 {\rm K})$. However, \citet{2011A&A...526A.156M} pointed out that the result might be biased toward larger values of $\dot{M}$ because the samples contained mainly extremely dusty RSG stars and the limitations of the mid-IR data available then.

Circumstellar dust forms with stellar mass loss and most RSGs have some amount of circumstellar dust \citep{2009A&A...498..127V}. These dust have obvious infrared emission, especially in the mid-infrared band, which makes the luminosity of such RSGs in infrared band exceed the upper limit of the theoretical luminosity of RSGs, and the contribution can be observed in the Spitzer 8$\mu$m and 24$\mu$m band.

RSGs are critical and important as both direct and indirect progenitors of supernovae that spend some time in the RSG phase. Lower mass RSGs are known to be the direct progenitors of core-collapse Type
\uppercase\expandafter{\romannumeral2}-P supernovae with clear hydrogen lines in optical spectrum and a distinctive plateau in visual lightcurves. A certain number of Type \uppercase\expandafter{\romannumeral2}-P supernovae have been proved to associate with RSG progenitors by observation \citep{2015PASA...32...16S, 2017MNRAS.469.2202M}. Higher mass RSGs, however, first noticed by \citet{2009MNRAS.395.1409S}, have not been detected as supernova progenitors. It may indicate that higher mass RSGs are not the direct progenitors of supernova, instead, they evolve back across the H-R diagram, spending a short time as yellow supergiants, blue supergiants or Wolf-Rayet stars before exploding as a supernova \citep{2012A&A...537..146E}. But high mass RSGs play an important role in their final fate because the mass-loss processes and dust production in the RSG phase will influence their ending.

Large amount of dust is detected in high redshift galaxies when the low-mass stars have not evolved to the AGB phase to become the dust contributor \citep{2013Natur.496..329R}. Therefore, the massive stars are the main if not unique providers of dust then. RSGs and supernovae (SNe) must be the major contributors to the interstellar dust in such distant galaxies.

Most RSGs show some degree of variability in the visual band. The typical amplitude of variation in the $V$ band is about 1 mag, while in the near-infrared the amplitude is smaller, about 0.25 mag in the $K$ band, and even smaller in the mid-infrared band \citep{2007ApJ...667..202L,2018A&A...616A.175Y}. According to the characteristics of light variation, the RSG variability is divided into two categories: one is irregular variation which is too complex to be delineated by any period; the other is semi-regular variation that can be further divided into two subclasses: one with short period ($\le$ several hundred days) and the other with Long Secondary Period (LSP) ($\ge$ thousand days). The main difficulty in the study of variability of RSGs is the long time scale of variation. The time-series data need to span over several hundreds of days to cover at least one entire period of variation. If the LSP is present and desired to be determined, an even longer time-baseline is required to last for several years.

Like other types of pulsating variables, RSGs exhibit the period-luminosity (P-L) relation. Both theory and observation prove the existence of the P-L relation in RSGs with semi-regular variation, especially the RSGs with short period variation. \citet{2002ApJ...565..559G} derived the theoretical P-L relations of fundamental (FU), first overtone (1O) and second overtone pulsation models for 15-30$M_\odot$ RSGs. In the Milky Way, based on the light curves collected by the American Association of Variable Star Observers (AAVSO) with a span of 60 years, \citet{2006MNRAS.372.1721K} studied 18 red supergiant stars to obtain their P-L relation.

Based mainly on the All Sky Automated Survey (ASAS) \citep{2002AcA....52..397P} and MAssive Compact Halo Objects (MACHO) \citep{1997ApJ...486..697A} observation, \citet{2011ApJ...727...53Y, 2012ApJ...754...35Y} analyzed the lightcurves of 112 RSGs in Small Magellanic Cloud (SMC) and 169 RSGs in LMC and determined the periods of those semi-regular variables. They derived the P-L relation based on 47 RSGs in LMC and 21 RSGs in SMC with short period semi-regular variation which was consistent with the result of theory \citep{2002ApJ...565..559G}. They also analyzed a sample of 40 RSGs in M33 and obtained their P-L relation in the 2MASS/$K_{\rm S}$ band by using the period from \citet{1987AJ.....93..833K}.

An extension to the environments of various metallicity will examine its effect on the P-L relationship. In the Local Group, the metallicity has a wide range from sub-solar in SMC (${\rm 12+ \log (O/H)=8.13}$; \citealt{1990ApJS...74..93R}) and LMC (${\rm 12+\log (O/H)=8.37}$; \citealt{1990ApJS...74..93R}) to solar in the Milky Way (${\rm 12+\log (O/H)=8.70}$; \citealt{1995RMxAC...3..133E}) and M33 (${\rm 12+ \log (O/H)=8.75}$; \citealt{1997ApJ...489..63G}) and then to super-solar in M31 (${\rm 12+ \log (O/H)=9.00}$; \citealt{1994ApJ...420..87Z}). It is known that the metallicity affects the ratios of blue-to-red supergiants (B/R) and Wolf-Rayet stars to RSGs (W-R/RSG). The B/R ratio changes by about 7 times \citep{1980A&A....90L..17M} and the W-R/RSG ratio changes by about 100 times over 0.9 dex in metallicity \citep{2002ApJS...141..81M}. Unfortunately, differences in W-R/RSG ratio cannot be fully explained by stellar evolutionary models of massive stars. Besides, metallicity has an obvious impact on color indices (or spectral types) and effective temperatures of RSGs. The spectral types of RSGs shift toward earlier at lower metallicities \citep{1985ApJS...57..91E, 2003AJ...126..2867M, 2012AJ...144..2L, 2016A&A...592..16D}. The effective temperatures of RSGs shift to warmer at lower metallicities, which was demonstrated by observation in different regions of M33 \citep{2012ApJ...750..97D}. The effect of metallicity on the P-L relation of RSGs is however not yet clear. From the studies of the variability of RSGs in SMC, LMC, the Milky Way, the P-L relations seem to be very similar despite of various metallicity. From the evolution and pulsation model of RSGs, for a given period, the higher the metallicity is, the fainter the luminosity is. The luminosity increases by 0.25 mag while metal abundance is doubled at a given period. On the other hand, the P-L relations at different metallicities show a tendency of increasing period with increasing metal abundance \citep{2002ApJ...565..559G}.

Now, the Intermediate Palomar Transient Factory (iPTF) survey \citep{2009PASP..121.1395L,2009PASP..121.1334R} covers the M33 and M31 sky area, and its time baseline is about 2000 days, providing a possibility for an elaborate analysis of the variation of RSGs in M33 and M31. \citet{2018ApJ...859...73S} already analyzed the P-L relation of RSG in M31 using iPTF data spanning from MJD 56000 to 58000. Here we analyze the case of M33. For consistency, M31 data is also included and checked the agreement with the result of \citet{2018ApJ...859...73S}.

\section{The sample of RSGs} \label{sec:style}

\subsection{Foreground stars}

A complete and pure sample of RSGs forms the solid basis to obtain a reliable P-L relation. Spectroscopy is the best tool to identify RSG in the Milky Way. Unfortunately, the extra-galactic RSGs may not be bright enough for high signal-to-noise ratio spectrum. On the other hand, they benefit the advantage of being at almost the same distance so that the apparent magnitude plays a role like absolute magnitude. Therefore, RSGs in the extragalactic systems can be identified by multiband photometry (e.g. \citealt{2006AJ....131.2478M}). The threshold of the luminosity and the range of color index are used to select the RSG candidates. For example, the $V = 20$ mag cut-off and color index $V- R \ge 0.85$ are used to select the RSG candidates in M31 \citep{2009ApJ...703..420M, 2016ApJ...826..224M}, where the $V = 20$ mag cut-off ensures sufficient brightness to block out the AGBs and the color index $V- R \ge 0.85$ restrict the stars to  K- and later type. The sample selected by this method can still be contaminated by foreground dwarf stars and red giants in the Galactic halo. \citet{1998ApJ...501..153M} proved that the combination of color indexes $V - R$ and $B - V$ can effectively separate RSGs from the foreground dwarfs, because $V - R$ is only sensitive to the effective temperature, while $B - V$ is sensitive to both the effective temperature and surface gravity so that the low-surface gravity RSGs and the high-surface gravity dwarfs are significantly different in $B - V$. Comparison of the Besan\c{c}on model \citep{2003A&A...409..523R} with the observed color-magnitude diagram (CMD) \citep{2016AJ....152...62M} and the kinetic identification (comparing the radial velocities of RSGs \citep{2009ApJ...703..420M} with the H\uppercase\expandafter{\romannumeral2} regions in M31 \citep{1970ApJ...159..379R}) found that the pollution probability of red giants in the Galactic halo was rather small. 
%

\begin{figure}[ht]
	\centering
	\includegraphics[width=0.7\textwidth]{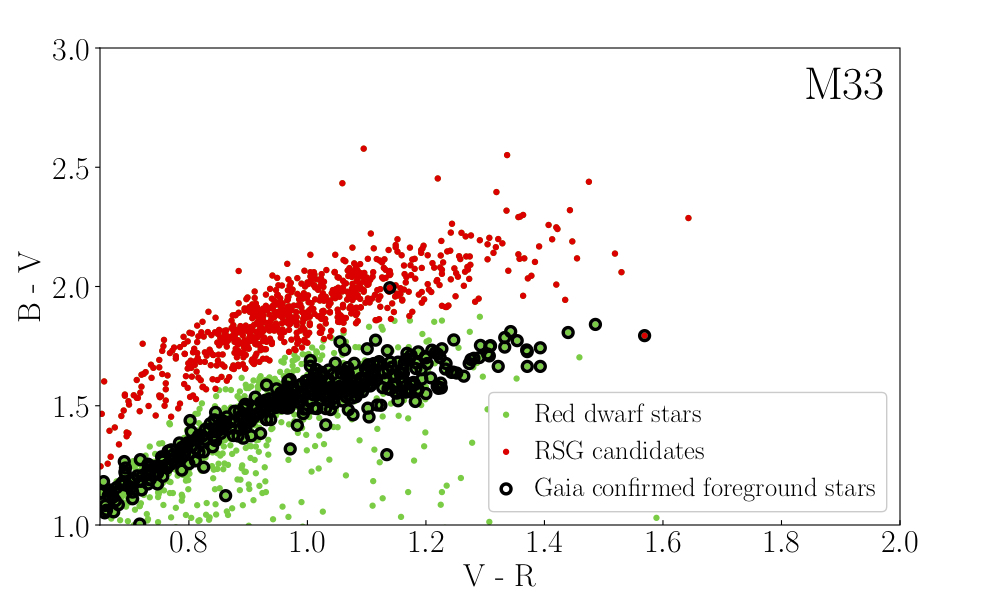}
	\caption{$B - V/V - R$ TCD of all red stars in M33. Foreground red dwarfs are marked as green dots, RSG candidates are marked as red dots, foreground stars confirmed by Gaia are marked as black circles.}
	\label{fig:Gaia_M33}
	
	\centering
	\includegraphics[width=0.7\textwidth]{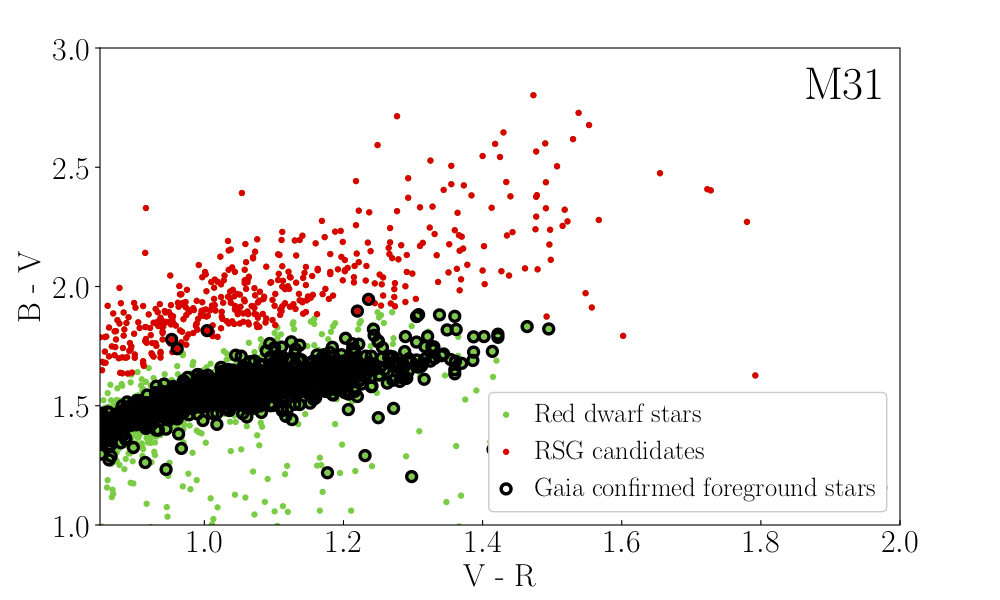}
	\caption{$B - V/V - R$ TCD of all red stars in M31. Foreground red dwarfs are marked as green dots, RSG candidates are marked as red dots, foreground stars confirmed by Gaia are marked as black circles.}
	\label{fig:Gaia_M31}
\end{figure}

The Local Group Galaxies Survey (LGGS) \citep{2006AJ....131.2478M, 2007AJ....133..2393M}
took images of several galaxies in the Local Group including M33 and M31 by using the 4-meter telescopes at the Kitt Peak National Observatory (KPNO) and the Cerro Tololo Inter-American Observatory (CTIO). This survey worked in multiple bands ($U, B, V, R, I$) at a 1$\sim$2\% photometry precision with a limiting magnitude of $\sim$21 mag. \citet{2009ApJ...703..420M} and \citet{2016ApJ...826..224M} published a sample of 437 RSG candidates in M31 based on the Two-Color Diagrams (TCD) method. We select RSG candidates in M33 by the same method with the criteria of $V < 20$, $V - R > 0.65$ and $B - V > -1.599(V - R)^2 + 4.18(V - R) - 0.83$ (see \citealt{2012ApJ...750..97D}) and obtain a sample of 749 RSG candidates in Fig.\ref{fig:Gaia_M33}. It can be seen that significant differences show up between supergiant candidates and foreground dwarfs which locate above and below the separating line respectively.

Although \citet{2009ApJ...703..420M,2016AJ....152...62M} demonstrate that the pollution probability of foreground red giants is small, we further use the distance of stars to check this argument directly with the help of the Gaia DR2 that provides parallaxes of 1.3 billion stars  \citep{2018A&A...616A...1G}. All the stars shown in Fig.\ref{fig:Gaia_M33} and Fig.\ref{fig:Gaia_M31}  are cross-matched with the Gaia DR2 and their distances are calculated with the Smith-Eichhorn correction method \citep{1996MNRAS.281..211S} from the GAIA-measured parallax. The distances of 918 red stars in M33 and 2216 red stars in M31 have relative error better than 20\%. Two (five) RSG candidates in M33 (M31) selected by the TCD method are found to have a distance modulus less than 24.66 (24.40) - the distance modulus of M33 (M31), which are marked by black circles in Fig.\ref{fig:Gaia_M33} and Fig.\ref{fig:Gaia_M31}. This confirmed that the pollution probability of red giants is truly small, on the scale of about one percent. Besides, foreground stars from TCD method are analysed to check the agreement with GAIA distances: 430 of 1677 foreground stars in M33 and 1161 of 3589 foreground stars in M31 are confirmed by the GAIA distances. Removing the foreground red giants from the TCD-selected samples, we are left with the samples of 747 RSG candidates in M33 and 432 RSG candidates in M31 consequently.

\subsection{The AGB stars}

\begin{figure}[ht]
	\centering
	\includegraphics[width=\textwidth]{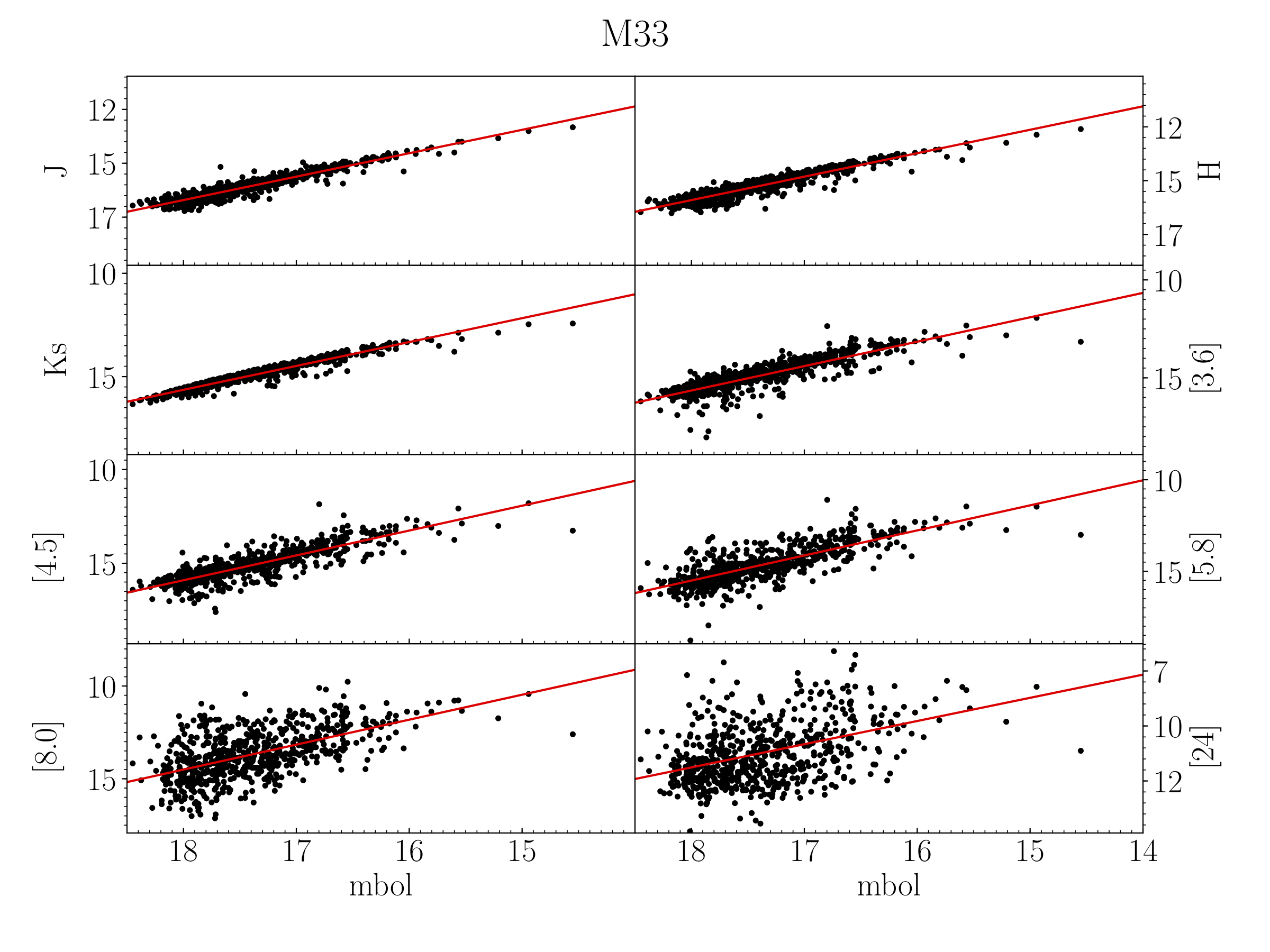}
	\caption{Linear fit between bolometric magnitude and multiband magnitude from RSG candidates in M33.}
	\label{fig:M33_mutilbandfit}	
\end{figure}

\begin{figure}[ht]	
	\centering
	\includegraphics[width=\textwidth]{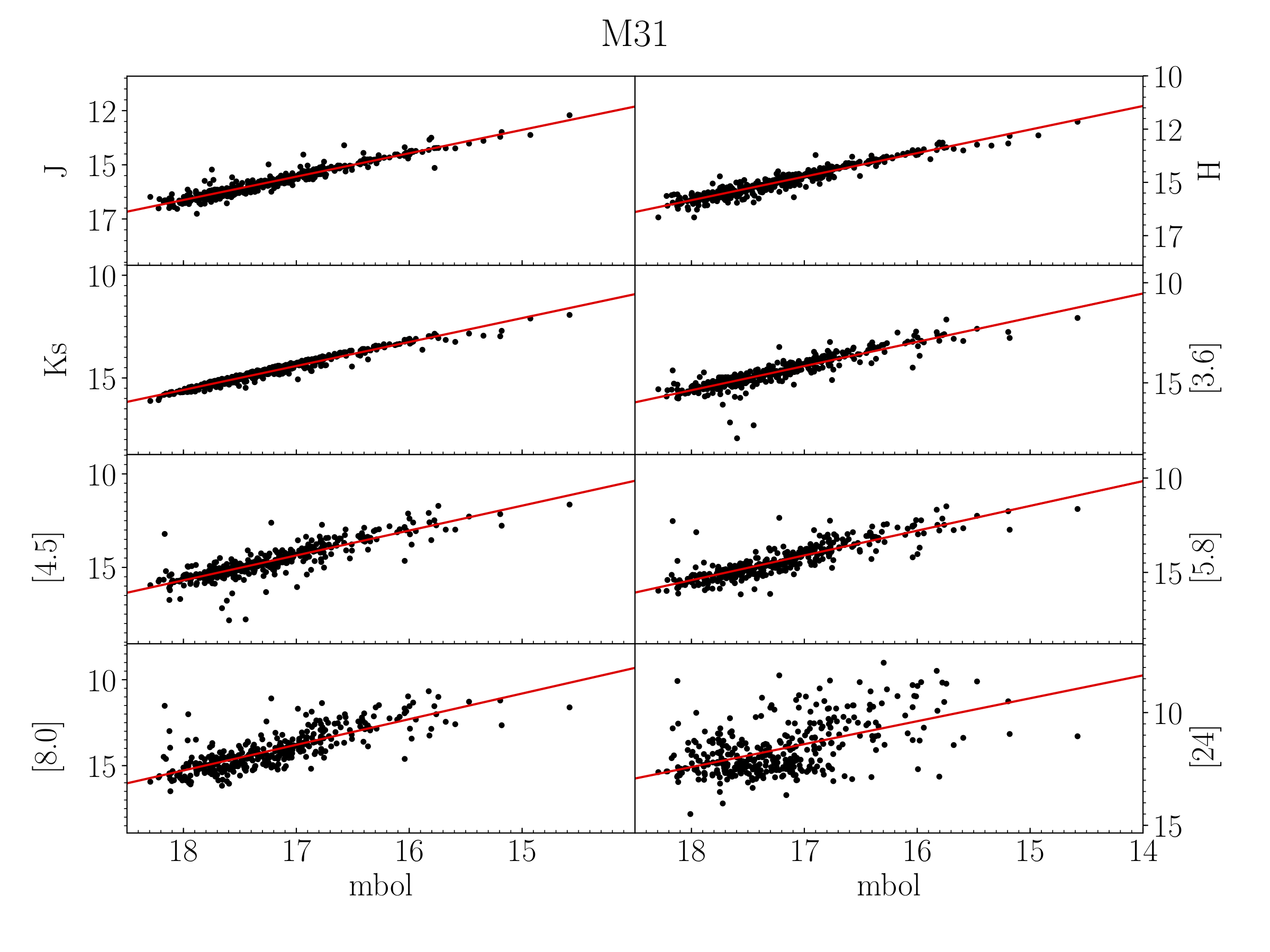}
	\caption{Linear fit between bolometric magnitude and multiband magnitude from RSG candidates in M31.}
	\label{fig:M31_mutilbandfit}
\end{figure}

Once the foreground stars are removed, the only pollution is the AGB stars in M33 and M31. On the HR diagram, the high luminosity AGB stars have a definite overlap with the low luminosity RSGs \citep{1986AJ...91...598B}. We solve this problem by setting the luminosity threshold in multiple bands. According to the mass-luminosity relation of massive stars:
$$ \frac{L}{L_\odot}  = (\frac{M}{M_\odot})^\gamma$$
where \begin{math}\gamma\approx4\end{math} \citep{1971A&A....10..290S}, the RSG luminosity corresponding to the mass range of 9-27${M_\odot}$  is ${9^4 - 27^4 L_\odot}$. With the relationship between luminosity and absolute bolometric magnitude:
$$M_{\rm bol} = 4.74-2.5\times\log(\frac{L}{{L_\odot}})$$
the upper and lower limits of $M_{\rm bol}$ are $M_{\rm bol}^{\max} = -9.57$ and $M_{\rm bol}^{\min} = -4.80$. Given the distance modulus of M33 being 24.66 \citep{2007Natur.449..872O} and of M31 being 24.40 \citep{2009A&A...507.1375P}, the limits of apparent bolometric magnitudes become $m_{\rm bol}^{\max} = 15.09$ and $m_{\rm bol}^{\min} = 19.86$ for M33, and $m_{\rm bol}^{\max} = 14.83$ and $m_{\rm bol}^{\min} = 19.60$ for M31. On the observational side, the bolometric $K$ correction in the $K$ band from \citet{1998A&A...333..231B,1998A&A...337..321B} is added to obtain the apparent bolometric magnitude from the apparent $K$ band for the RSG stars within given mass range. First, the effective temperature is derived from the intrinsic color index $(V - K)_{0}$: $T_{\rm eff} = 8130.9 - 2113.22(V - K)_{0} + 327.883(V - K)^2_{0} - 17.7886(V - K)^3_{0}$, where $(V - K)_{0}$ is calculated from the observed color by subtracting the interstellar extinction by assuming $A_{V} = 1$ and $A_{K}/A_{V} = 0.12$ for all RSG candidates \citep{2016ApJ...826..224M}. Then, the bolometric correction in the K band is calculated from ${\rm BC}_{K} = 7.149 - 1.5924(T_{\rm eff}/1000{\rm K}) + 0.10956(T_{\rm eff}/1000{\rm K})^2$ \citep{1998A&A...333..231B,1998A&A...337..321B} where the difference between the $K$ and $K_{\rm S}$ bands is very small and ignored in this work. The linear relationship of the apparent magnitude with the apparent bolometric magnitude is derived, which is used to convert the limits of bolometric magnitudes to that in the corresponding band. The results of linear fitting between bolometric magnitude and that in the $\lambda$ band are shown in Fig.\ref{fig:M33_mutilbandfit} and \ref{fig:M31_mutilbandfit}, e.g. the upper and lower limits in the $J$ band is derived to be 13.54 mag and 18.73 mag which correspond to the limits of apparent bolometric magnitudes of $m_{\rm bol}^{\max} = 15.09$ and $m_{\rm bol}^{\min} = 19.86$ for M33. The case of M31 is presented in Table \ref{tab:M31_mutilbandfit} and Fig.\ref{fig:M31_mutilbandfit}.


It is worthy to note that the linear fitting has relatively large dispersion in the Spitzer 8$\mu$m and 24$\mu$m bands. This can be understood by the contribution of circumstellar dust to the flux at 8$\mu$m and 24$\mu$m in addition to the stellar photosphere. This scattering leads to relatively uncertain limits of brightness in the 8$\mu$m and 24$\mu$m bands.

\begin{figure}[ht]	
	\centering
	\includegraphics[width=\textwidth]{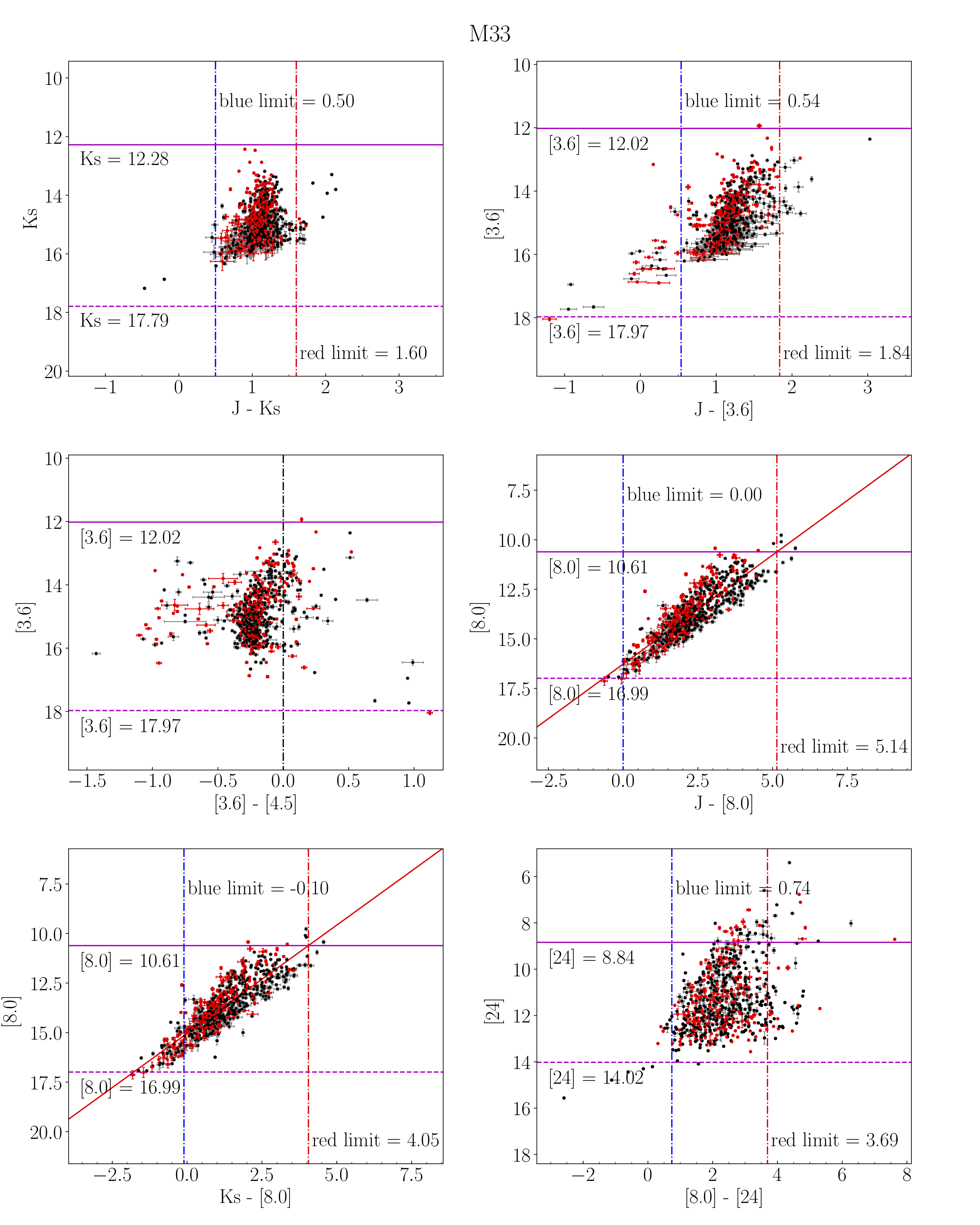}
	\caption{CMDs for RSG candidates in M33. Magenta solid line indicates uppper limit of magnitude in each band, Magenta dash line indicates lower limit of magnitude in each band, red dashdotted line indicates red limit of RSGs, blue dashdotted line indicates blue limit of RSGs. Red dots are sources confirmed by spectrum.}
	\label{fig:M33_CMD}
\end{figure}

\begin{figure}[ht]	
	\centering
	\includegraphics[width=\textwidth]{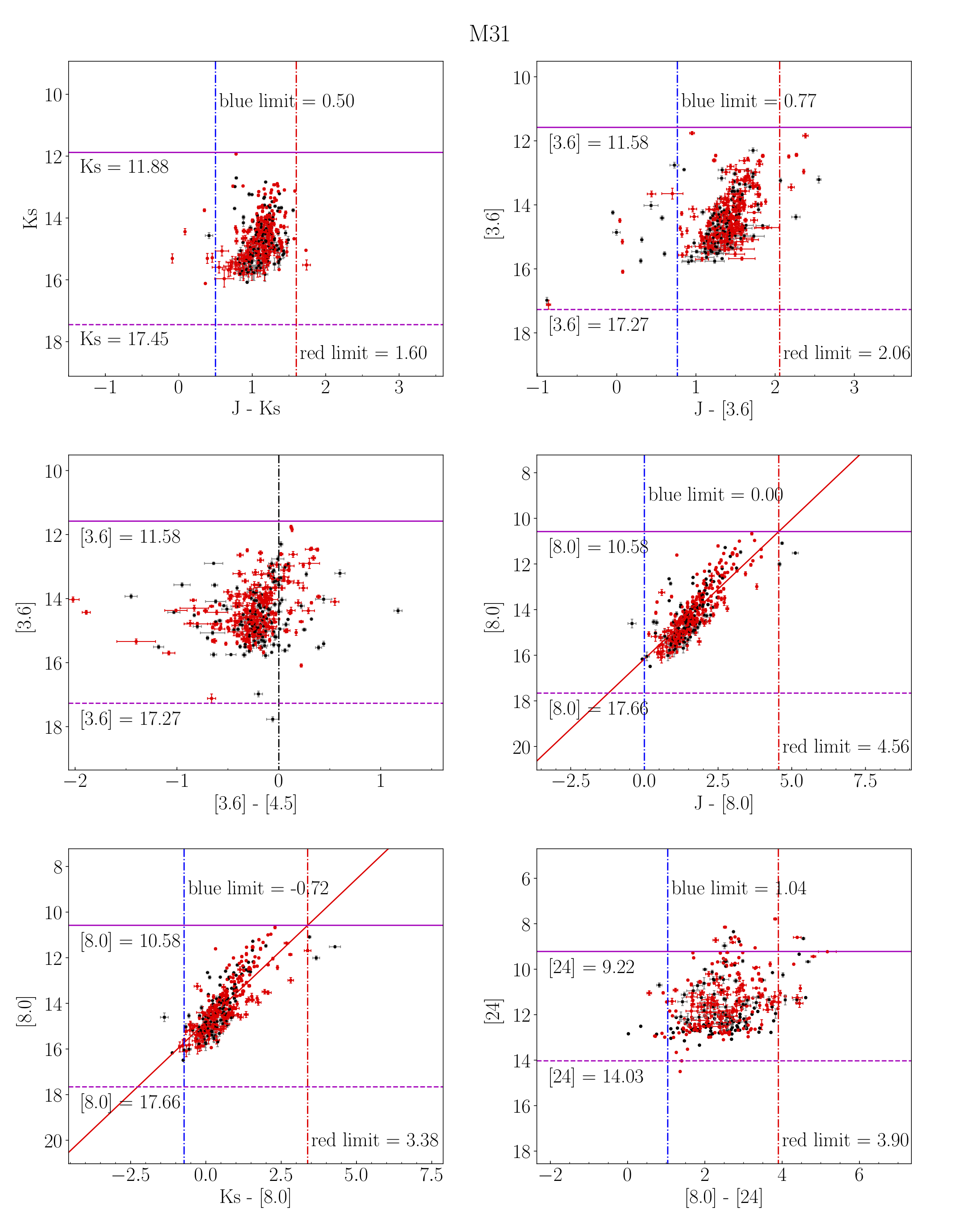}
	\caption{CMDs for RSG candidates in M31. The symbol convention is the same as Fig. \ref{fig:M33_CMD}.}
	\label{fig:M31_CMD}
\end{figure}

Once the range of the brightness in some bands, mainly the lower limit of the brightness, is set up, the locations of the RSG stars can be delineated in the CMD with the help of the color index and the AGB stars can be separated.  Several near- and mid-infrared CMDs are shown in Fig.\ref{fig:M33_CMD} and Fig.\ref{fig:M31_CMD}, where the magenta solid line marks the upper limit of luminosity and the magenta dashed line marks the lower limit of luminosity. The range of color index is marked by red dashdot line and blue dashdot line. From multiple CMDs of M31 and M33, the following facts are observed:
\begin{enumerate}
  \item  The maximum luminosity of all sources agrees well with the theoretical limit of RSGs corresponding to $M= 27{M_{\odot}}$.
  \item  The minimum luminosity of all sources is about one magnitude above the theoretical limit of RSGs corresponding to 9$M_{\odot}$. The reason may be the selection effect of observation that the fainter RSGs are not detected due to the sensitivity limit.
  \item  In the CMD of $K_{\rm S}$ vs $J - K_{\rm S}$, the red limit is at $J - K_{\rm S} = 1.60$, which is the limit of the carbon-rich and oxygen-rich stars from \citet{1990AJ.....99..784H}. The blue limit is at $J - K_{\rm S} = 0.50$, which is the observational limit of the RSGs in the Galactic halo obtained by \citet{2000A&A...357..225J}.
  \item  In the  [3.6] vs $J-[3.6]$ diagram, about 95\% of the color index values lie within two standard deviations, the limits of blue and red colors are set to include over 95\% sources in the sample based on the distribution.
  \item  Because of the influence of the CO absorption line on the 4.5$\mu$m band, the RSGs appear fainter in this band. So they look blue in the color index [3.6]-[4.5] and most of the sources have negative values.
  \item  In the CMDs of [8.0] vs $J - [8.0]$ and [8.0] vs $K_{\rm S} - [8.0]$, $K_{\rm S} - [8.0]$ becomes redder along with the increasing brightness in the [8.0] band due to the emission of circumstellar dust. There even exists a good linear relationship between the color index and the brightness at [8.0].
            For M33, $[8.0] = -1.10 \times (J - [8.0]) + 16.26$ and $[8.0] = -1.09 \times (K_{\rm S} - [8.0]) + 15.03$.
            For M31, $[8.0] = -1.20 \times (J - [8.0]) + 16.11$ and $[8.0] = -1.24 \times (K_{\rm S} - [8.0]) + 14.80$.
            Therefore, the red limit of the color index is taken at the intersection point of the fitting line and the upper limit of the luminosity. Once the red limit is set, the blue limit is set to ensure the probability that lies in the range between the limits of blue and red colors is 95\% based on the distribution.
  \item  In the CMD of [24] vs [8.0] - [24], due to the influence of the infrared emission of circumstellar dust on the 24$\mu$m band, some sources supersede the theoretical luminosity upper limit of the RSGs. The limits of the color index [8.0] - [24] are set to include over 95\% stars in the sample based on the distribution.
\end{enumerate}

In order to clear of the pollution of AGB stars, sources which are fainter than the lower limit of $K_{\rm S}$, $[3.6]$, $[8.0]$, $[24]$ bands are removed.

Sources which are bluer than the blue limit of $J - K_{\rm S}$ or the blue limit of $J - [3.6]$ should be removed as well. As for color indexes of $J - [8.0]$, $K_{\rm S} - [8.0]$, $[8.0] - [24]$, they are not reasonable criterion due to the impact of circumstellar dust, we only mark by experience the range of these three color index.

In the CMDs of M33, 36 sources are bluer than the blue limit of $J - K_{\rm S}$ or $J - [3.6]$, however, 16 of them have been identified as RSGs by spectroscopy \citep{2016ApJ...826..224M}. In addition, 1, 1, 2, 6 sources are fainter than the lower limit of the Spitzer/IRAC [3.6], [5.8], [8.0] and Spitzer/MIPS [24] band respectively. In total, 30 sources are removed finally.

For M31, 22 sources are bluer than the blue limit of $J - K_{\rm S}$ or $J - [3.6]$, however, 11 of them have been identified as K-type or M-type supergiant stars by spectroscopy \citep{2016ApJ...826..224M} so that they are kept, and 1 of these sources is also fainter than the lower limit of Spitzer/IRAC [3.6] band magnitude. Besides, 1 source is fainter than the lower limit of the Spitzer/MIPS [24] band magnitude and removed. In total, 12 sources are removed.

We are left with the samples of 717 RSGs in M33 and 420 RSGs in M31 at last.

\section{Time-domain data} \label{sec:style}

\subsection{The iPTF data}
\begin{figure}[ht]	
	\centering
	\includegraphics[width=0.9\textwidth]{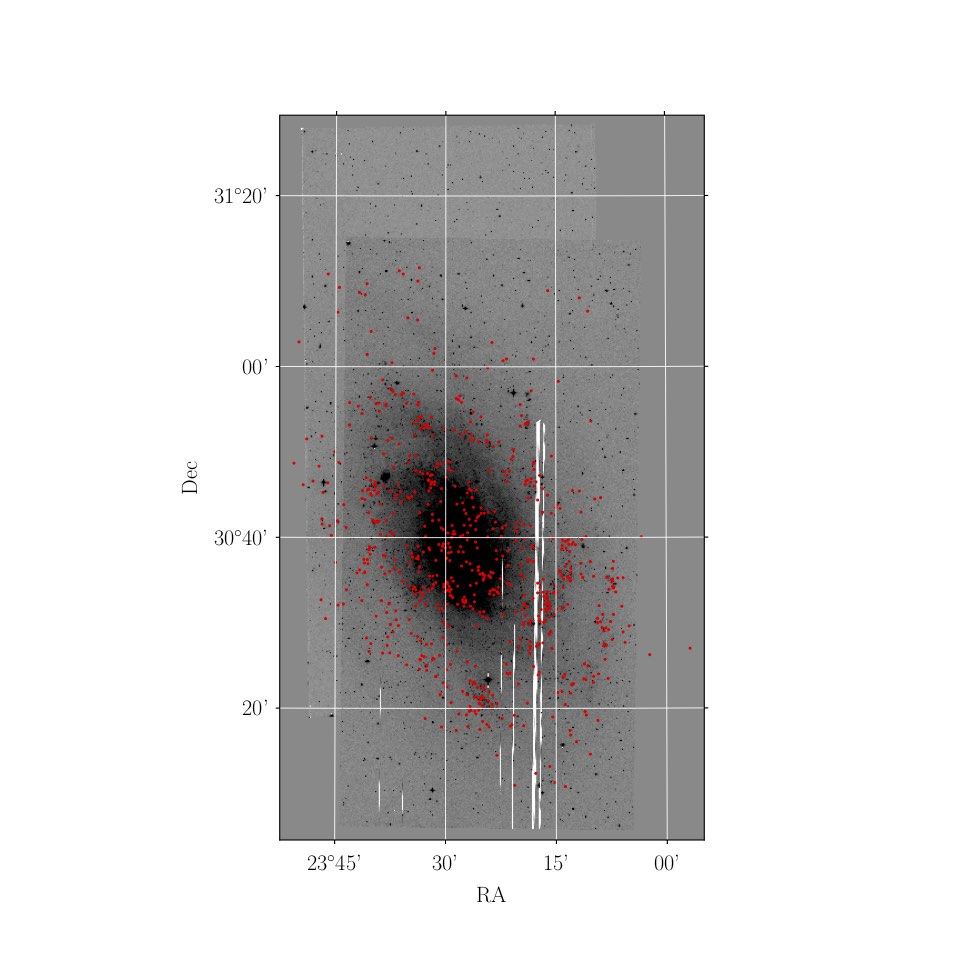}
	\caption{The iPTF image of M33 with the red dot for the position of RSGs.}
	\label{fig:image_M33}
\end{figure}

\begin{figure}[ht]	
	\centering
	\includegraphics[width=0.9\textwidth]{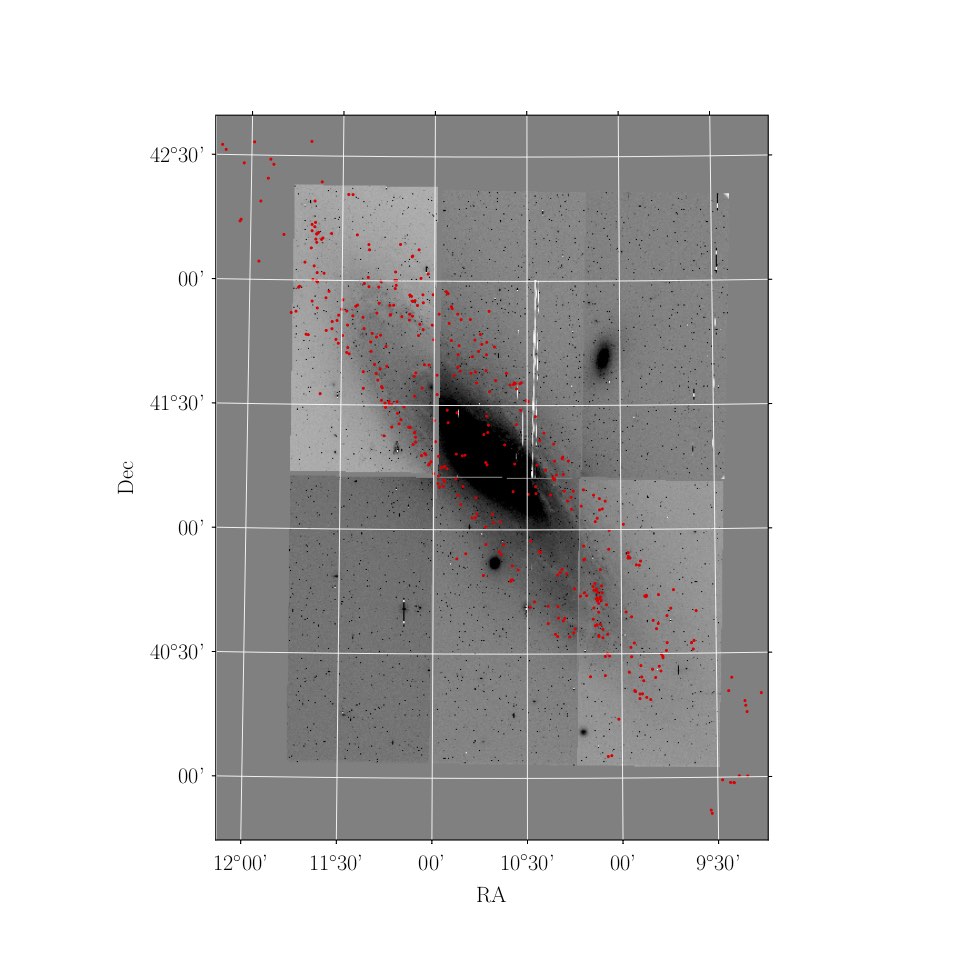}
	\caption{The iPTF image of M31 with the red dot for the position of RSGs.}
	\label{fig:image_M31}
\end{figure}

The time-series data is taken from the iPTF survey, a wide-field survey for an exploration of optical transients using the 1.2-meter Samuel Oschin Telescope at Palomar Observatory \citep{2009PASP..121.1395L,2009PASP..121.1334R}. This survey uses a large field camera covering 7.8 square degrees with 11 2048 $\times$ 4096 CCDs at a resolution of 1.01 arcsec/pixel.
In the iPTF survey, 2 and 6 frames covered the M33 and M31 sky area respectively from 2009 Aug to 2015 Jan (about 2000 days in total) with a typical seeing of 2 arcseconds. Although the images were taken in both $R$ and $g$ band, the $R$ band observations are dominant and make use of \textgreater 80\% time, reaching 20.5 mag at a 5-sigma level. Therefore we choose the $R$ band iPTF data to analyze the variation of RSGs.
The RSGs are selected in the iPTF images within three arcseconds from the position of the star. Fig.\ref{fig:image_M33} and Fig.\ref{fig:image_M31} show the iPTF images that include 658 and 377 RSG stars in M33 and M31 respectively, whose light variations are further analyzed.

\subsection{Photometry}

\begin{figure}[ht]
	\centering
	\includegraphics[width=0.85\textwidth]{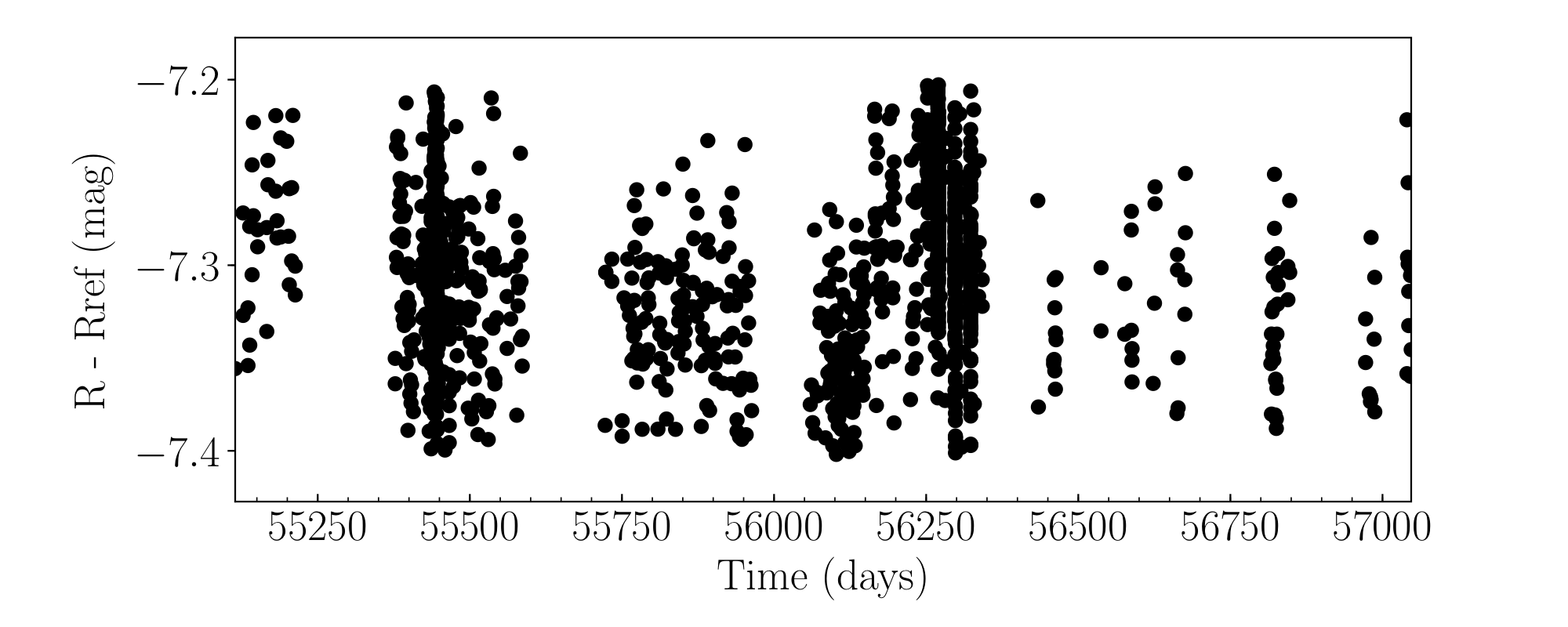}
	\caption{Performance of the artificial reference star. Rref represents $R$ band magnitude of the artificial reference star, $R$ represents $R$ band magnitude of the selected Tycho reference star.}
	\label{fig:ref_performance}
	
	\centering
	\includegraphics[width=\textwidth]{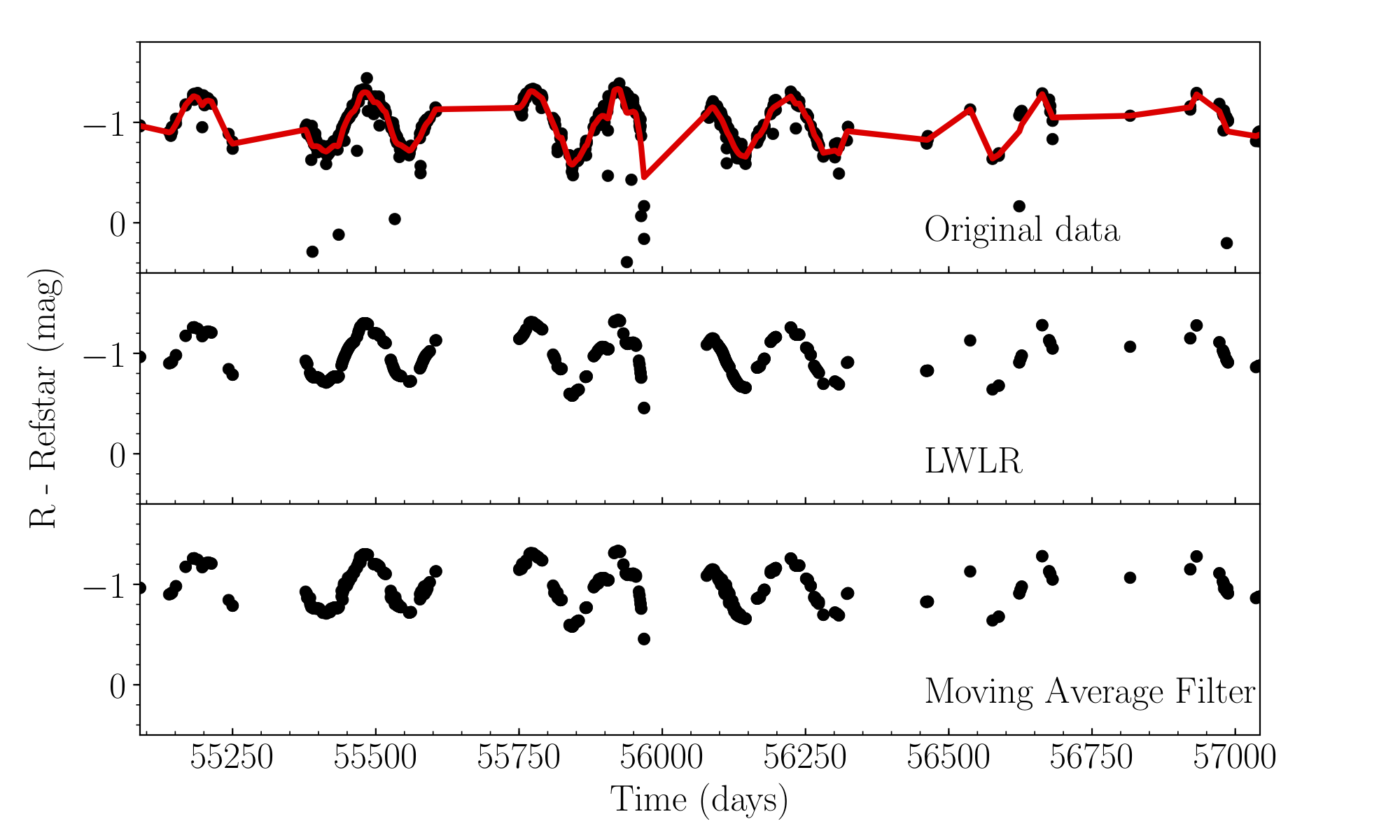}
	\caption{Data preprocessing. The top panel shows the original data, the red line shows the fitting curve using locally weighted linear regression. The middle panel shows the data after removing the points which lie more than 3$\sigma$ away from the lightcurve. The bottom panel shows the data that is smoothed by moving average filter.}
	\label{fig:smooth}
\end{figure}

The CCD images were processed in a standard way. We measured all the iPTF images available (796 images of M33, 1980 images of M31) using the Sextractor code \citep{1996A&AS..117..393B} with fixed coordinates. For a precise photometry, a stable reference star should be chosen for differential photometry, and its brightness and color should be similar to the target stars in order to cancel the influence of the instruments and atmosphere as much as possible. For this purpose, we made an artificial reference star whose brightness is the average of all the red supergiant stars in each image, i.e. the average of approximately 700 RSGs in M33 and 100 RSGs in M31 within one image. Naturally, this artificial star has the very typical color of RSGs. The large scale of the sample should smooth out the variation of all red supergiant stars, which makes the artificial reference star stable. The stability of the brightness is checked by comparing with a star from the Tycho reference catalogue \citep{1998A&A...335L..65H} with a brightness of $VT = 12.608$ mag and a color of $VT - BT = 1.704$ similar to RSGs. As shown in Fig.\ref{fig:ref_performance},  the standard error of differential photometry between the artificial reference star and the Tycho reference star is about 0.04 mag, which implies the photometric accuracy to be better than 0.04 mag. Considering this Tycho reference star is much brighter than the RSGs, the real accuracy should be even better. In order to achieve a good photometry quality, the size of the aperture is adjusted according to the full width at half-maximum (FWHM) of the PSF in each image by $R_{\rm aperture} = 1.5 + 1.2 \times \rm FWHM$.

\section{Period Determination}

Before we start to determine the period of light variation, the original photometric data from differential photometry are processed further.
First the outlying points are removed by using the locally weight linear regression (LWLR), with its most common method known as locally estimated scatterplot smoothing (LOESS). The weight function used for LOESS is the Gaussian function: $\omega(x,z) = {\rm exp}(-(x - z)^2/2\kappa^2)$ where we adopted $\kappa = 4.5$ with eye-check after trying the value in the range of [0.01,10]. The points lying more than 3$\sigma$ away from the locally fitting line are removed. Then the moving-average filter was used to smooth the light curve, in which a window of 9 days is taken. Fig.\ref{fig:smooth} illustrates an example showing the two processes and the resulted light curve for period determination.

For the period determination of non-uniformly sampled data as the present case, no single algorithm is fully suitable. It is necessary to use various methods to cross-check the existence of periodicity. The popular methods for determining period are the Phase Dispersion Minimization (PDM) \citep{1978ApJ...224..953S} algorithm based on the time domain, the Generalised Lomb-Scargle periodogram (GLS) \citep{2009A&A...496..577Z} algorithm based on the frequency domain, the Weighted Wavelet Z-transform (WWZ) \citep{1996AJ....112.1709F} algorithm based on the time-frequency and the discrete Fourier transform (DFT) based on frequency domain. All these four methods are applied to the light curves of the RSGs in M33 and M31.

The PDM method looks for the period that brings about the minimum phase dispersion designated by $\Theta$ in a number of trial periods. The GLS method is an improvement on the traditional Lomb-Scargle (LS) \citep{1976Ap&SS..39..447L, 1982ApJ...263..835S} method. The main improvement is that the GLS algorithm uses the sine function of the addition constant $\rm C$ to fit the time-series data and each calculation takes into account the effect of observation error $\omega$. Therefore, the period estimation accuracy of the GLS algorithm is better than that of the LS algorithm. The WWZ method is a period determination algorithm based on wavelet analysis and vector projection, very suitable for the analysis of non-stationary signals and has advantages in the analysis of time-frequency local characteristics. The DFT method is a classical period determination algorithm based on Fourier transform.

\begin{figure}[ht]	
	\centering
	\includegraphics[width=\textwidth,height=10cm]{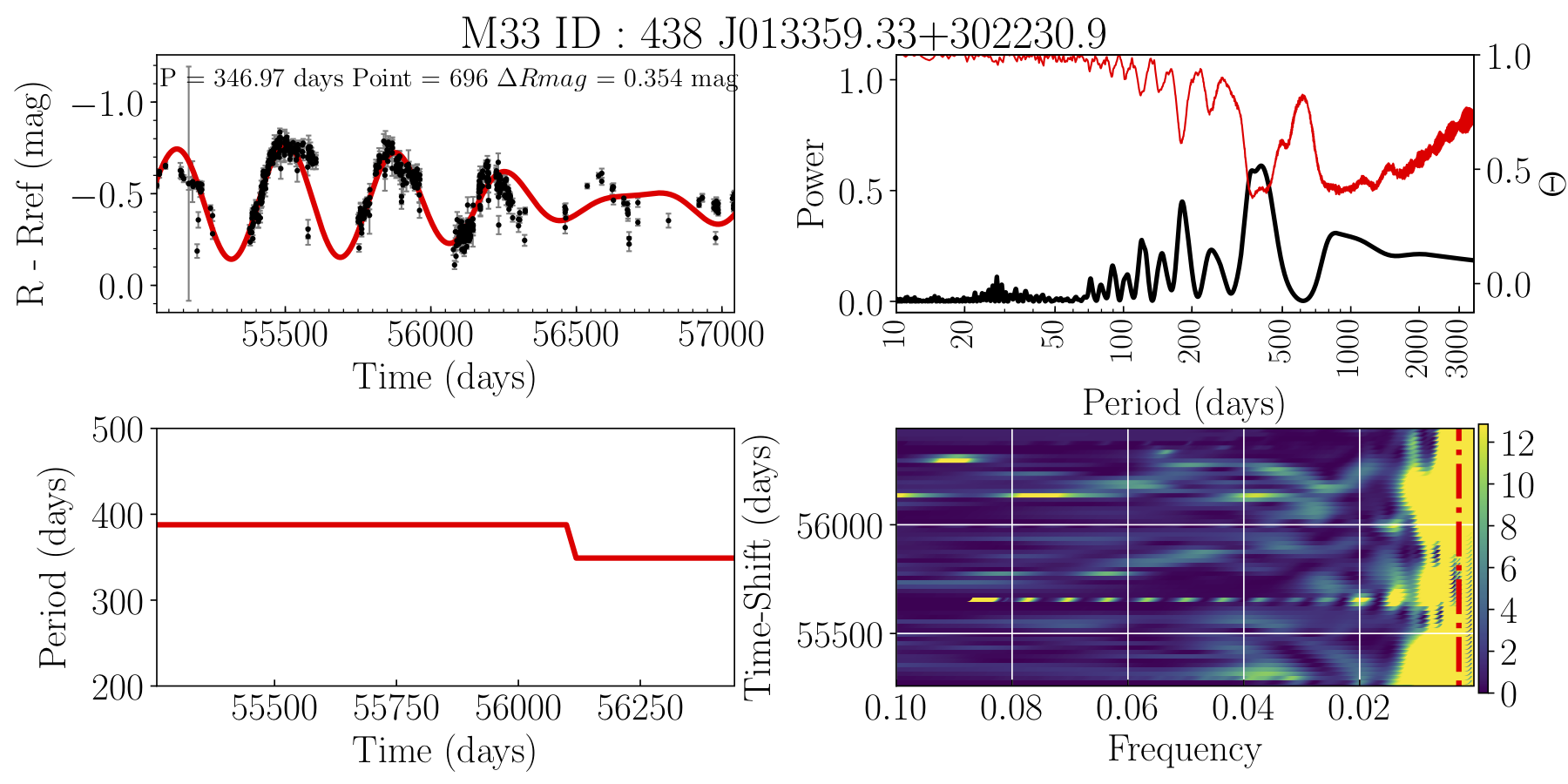}
	\caption{Example of the period determination from the lightcurve of a RSG with semi-regular variation.  The left top panel is the lightcurve (black dots with error bar) and the fitting result (red line).  The results of the PDM ($\Theta$ - P, red line) and GLS (Power - P, black line) analysis are shown in the top right panel. The result from WWZ is shown in the lower panels, where the right shows the WWZ map and the left shows the period against time.
	}
	\label{fig:example_period}
\end{figure}

We use the GLS, DFT and WWZ methods from VARTOOLS \citep{2016A&C....17....1H} and PDM \citep{1978ApJ...224..953S} method from PyAstronomy for the period determination.
The parameters used in the methods are shown in Table \ref{tab:The parameters used for the period determination} in details. The final selection of the period depends on the goodness of fitting the lightcurve. A period is assigned to an object with semi-regular variation only when it is detected by at least two methods whose difference in period is usually ($>70\%$) less than 40 days while up to 50-90 days in a few cases.
The period is searched continuously in the power spectrum of the residual data after subtracting the light variation with the most prominent period until the local signal to noise ratio of the power spectrum is less than 4. Fig.\ref{fig:example_period} shows an example of period determination, where both the PDM and GLS methods find a period at $\sim$ 347 days, and the WWZ method also confirms that this period is present in most of the observational time.
Based on the period and the regularity of light variation, these RSGs are classified into five categories:
\begin{enumerate}
	\item Semi-regular variables, marked as \textbf{S}, which have definitely a period and are used for the follow-up analysis of the P-L relations. There are 84 (56) sources in M33 (M31).
	
	\item LSP variables, marked as \textbf{L}, which show evidence of LSP while the LSP cannot be determined accurately due to the limited time span of the iPTF data ($\sim$ 2000 days). There are 48 (13) sources in M33 (M31).
	
	\item Irregular variables, marked as \textbf{I}, which have small amplitude and too complex variation to show any periodicity. There are 14 (21) sources in M33 (M31).
	
	\item Undefined class, marked as \textbf{U}, which lack enough observational data to determine its variability. There are 86 (18) sources in M33 (M31).
	
	\item No significant variables, marked as \textbf{N}, which have no significant variation. There are 426 (269) sources in M33 (M31).
\end{enumerate}

The light curves of the examples of the first three types are shown in Fig.\ref{fig:type}.

The light curves with the fitting of all stars are displayed in Appendix \ref{appendix A}, and the properties of variation and photometric results are appended in Appendix \ref{appendix B}.

\begin{figure}[ht]	
	\centering
	\includegraphics[width=\textwidth]{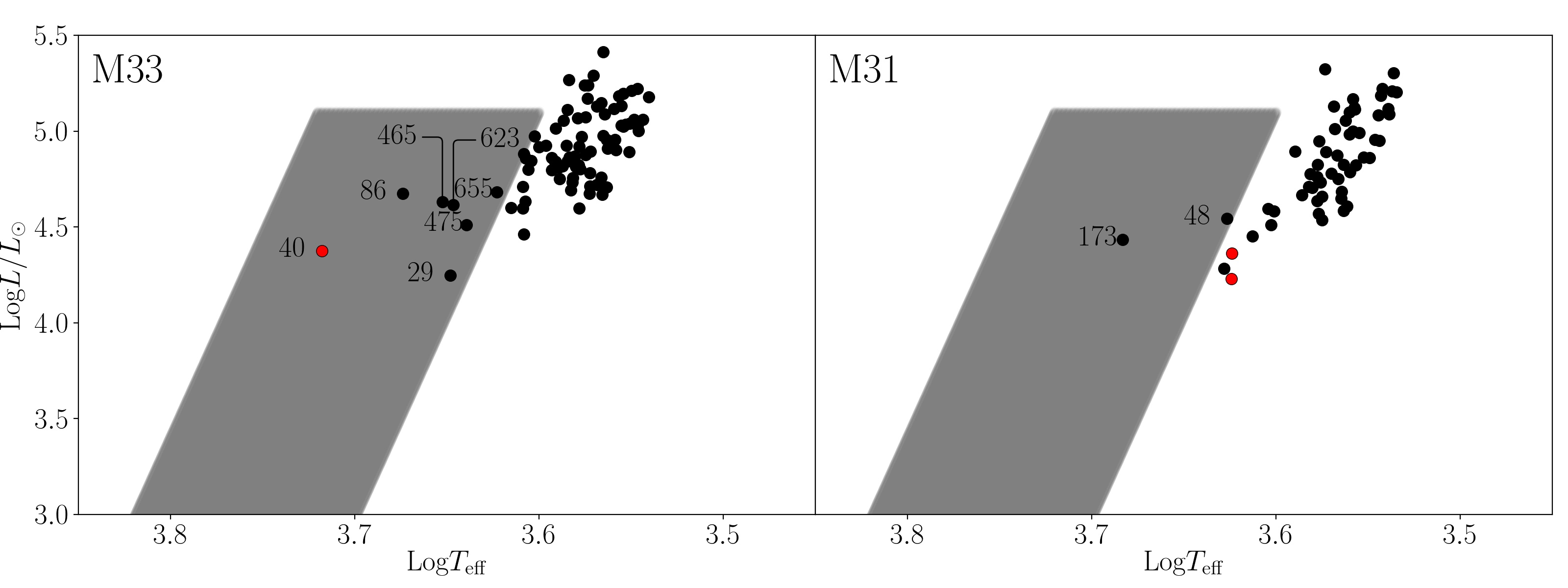}
	\caption{The Hertzsprung-Russell diagram for RSGs with semi-regular variation in M33 (left) and M31 (right). The grey shaded area represents the instability strip \citep{2003A&A...404..423T}. Red dots are the regular variables, and black dots with ID from Table \ref{tab:Parameters of RSGs in M33} and  \ref{tab:Parameters of RSGs in M31} are the RSGs in the instability strip.}
	\label{fig:IS}
\end{figure}

In addition, three stars exhibit very regular light variation, which are J004115.35+405025.2, J004415.42+413409.5 in M31 and J013237.47+301833.9 in M33. Among them, J004115.35+405025.2, J004415.42+413409.5 are also studied by \citet{2018ApJ...859...73S}. These sources have relatively short period (96.5, 76.5 and 146.5 days respectively). The spectral types are identified as K1 for J004115.35+405025.2, K2 for J004415.42+413409.5 and RSG for J004415.42+413409.5 \citep{2016ApJ...826..224M}. As discussed previously, the luminosity of these stars indicates that they are supergiants which means that some RSGs may have regular variation. It will be shown in the next section that these stars very possibly pulsate in the first overtone mode. These three RSGs are plotted in the Hertzsprung-Russell diagram (HRD) in Fig.\ref{fig:IS}. It can be seen that RSG J013237.47+301833.9 in M33 lies in the instability strip (IS), while J004115.35+405025.2, J004415.42+413409.5 in M31 are very close  to the IS.  Moreover, these stars are at the faint end of the RSGs, which implies relatively low-mass. We may say that these regular variables are RSGs with relatively low mass and relatively high temperature. On the other hand, there are some other sources also in the IS and they are not regular variables but semi-regular variables. It is known that yellow supergiant (YSG) is a kind of Cepheid, meanwhile RSGs and YSGs are hard to distinct near the blue limit of RSGs as the three regular variables are. It can be concluded that Some RSGs with early K spectral types may lie in the IS and have very regular variations.

\section{Period-Luminosity Relation}

The P-L relation is derived for the bands with photometry data available, from the optical V band through 2MASS near-infrared $JHK_{\rm S}$ bands to the Spitzer/IRAC and Spitzer/MIPS bands. As mentioned earlier, we only adopted the RSGs that show semi-regular variation and whose period are found by at least two methods.

\begin{figure}[ht]	
	\centering
	\includegraphics[width=\textwidth]{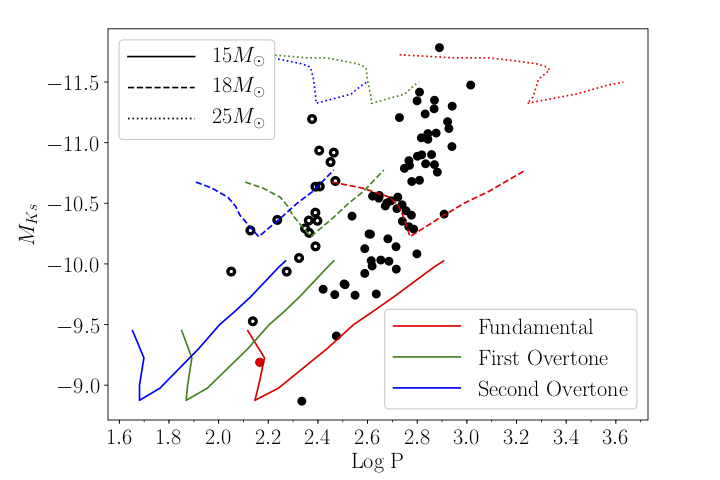}
	\caption{Locations of the RSGs in M33 in the absolute $K_{\rm S}$ magnitude vs. $\log \rm P$ diagram. The lines are based on the model calculations by \citet{2002ApJ...565..559G} at metallicity Z = 0.02. The solid black dots are identified as pulsating in the fundamental pulsation mode, the black circles in the first overtone pulsation mode, and the red dot is the RSG with regular variation.}
	\label{fig:P_L_model_M33_ALL}
\end{figure}

\begin{figure}[ht]	
	\centering
	\includegraphics[width=\textwidth]{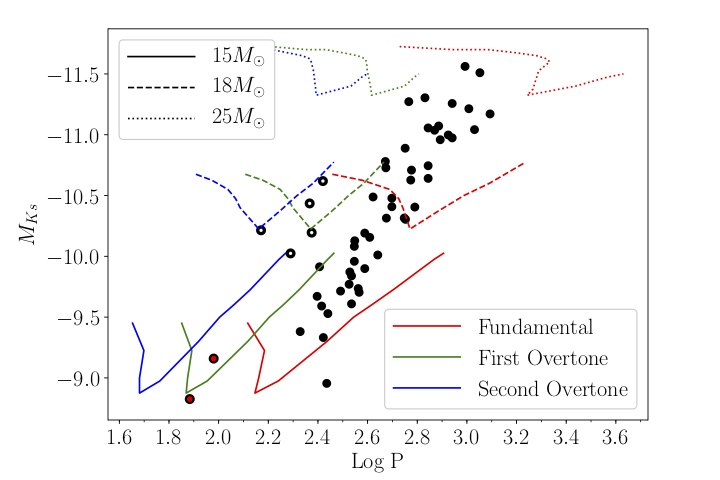}
	\caption{The same as Fig. \ref{fig:P_L_model_M33_ALL}, but for M31.}
	\label{fig:P_L_model_M31_ALL}
\end{figure}

The observational results are compared with the model by \citet{2002ApJ...565..559G} in Fig.\ref{fig:P_L_model_M33_ALL} and Fig.\ref{fig:P_L_model_M31_ALL}. The absolute magnitude in the $K_{\rm S}$ band of the model is obtained from the absolute bolometric luminosity with the help of the bolometric correction in the $K_{\rm S}$ band by \citet{2000A&A...357..225J}. Although the models are not perfectly matched with the observation, two sequences are clear in that most of the stars follow the sequence of the fundamental-mode pulsation and a few stars are closer to the first overtone pulsation mode. The three regular variables all locate close to the first overtone model. Because the pulsations of the fundamental and first overtone modes follow different P-L relations, they must be separated to study this relation. We identify 19 stars in M33 and 7 stars in M31 pulsating in the first overtone mode, and their P-L relation is not analyzed due to too few stars. Nevertheless, there is no clear indication of a P-L relation for them. The P-L relation of the 65 RSGs in M33 and 49 RSGs in M31 that pulsate in the fundamental mode is then analyzed.

%

In principle, the brightness should be the mean magnitude when the P-L relation is dealt with. However, the brightness in the $V$ band is taken from the single measurement by LGGS. Since the amplitude of variation in the $V$ band is $\sim 1$ mag, this can cause some dispersion in the P-L relation in the $V$ band. Although the brightness in the infrared bands is also taken from the single measurement, the small amplitude of variation makes the single measurement approximate to the mean brightness and do not bring significant scatter in the P-L relation.

\begin{figure}[ht]	
	\centering
	\includegraphics[width=\textwidth]{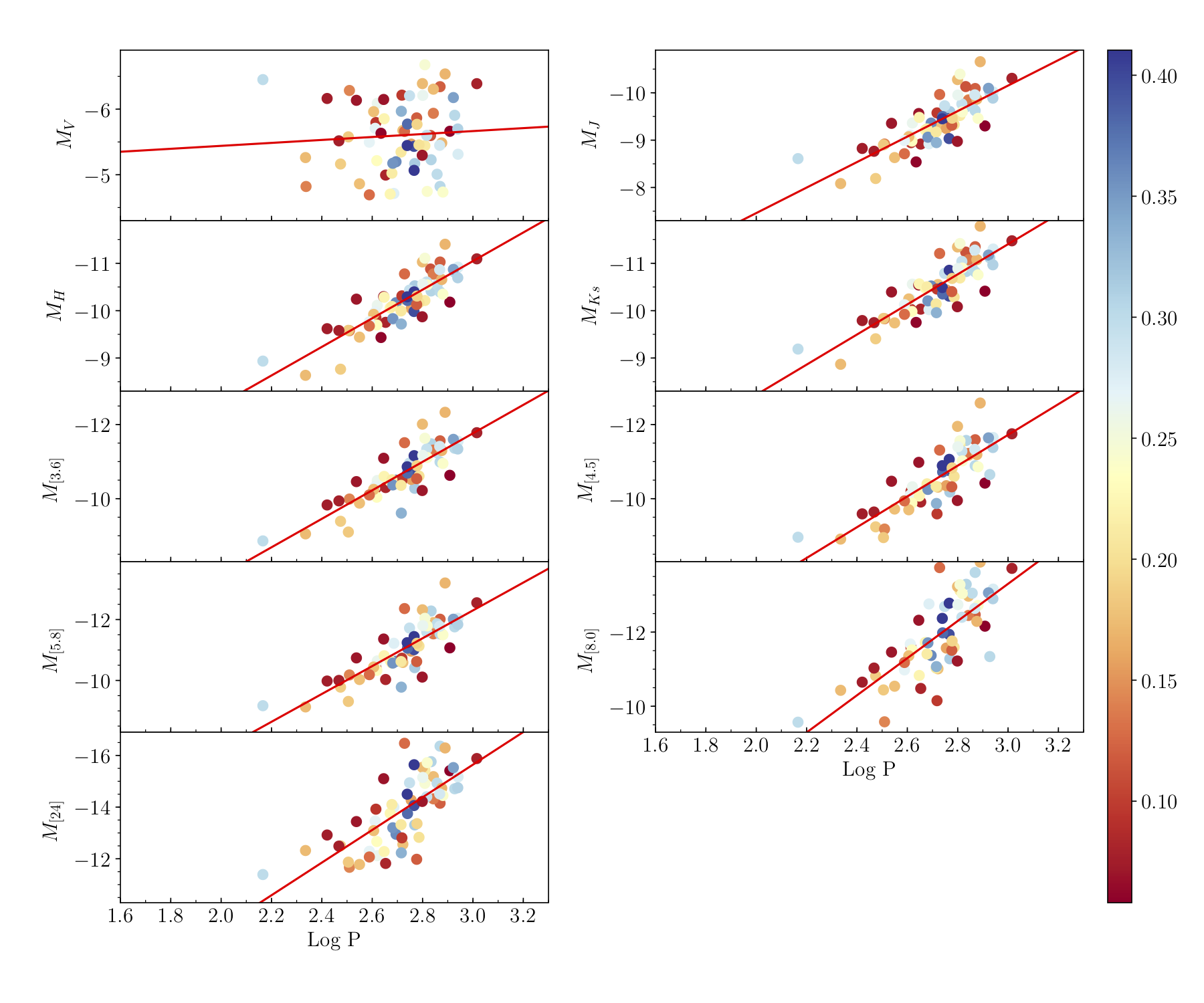}
	\caption{The Period-Luminosity relation in the $V, J, H, K_{\rm S}$, Spitzer/IRAC [3.6], [4.5], [5.8], [8.0] and MIPS [24] bands for RSGs with semi-regular variation in M33. The red solid line is a linear fit between the multiband absolute magnitude and the period. The color of dots decodes the variation amplitude in accord with the colorbar (in the unit of magnitude).}
	\label{fig:P_L_M33}
\end{figure}

\begin{figure}[ht]	
	\centering
	\includegraphics[width=\textwidth]{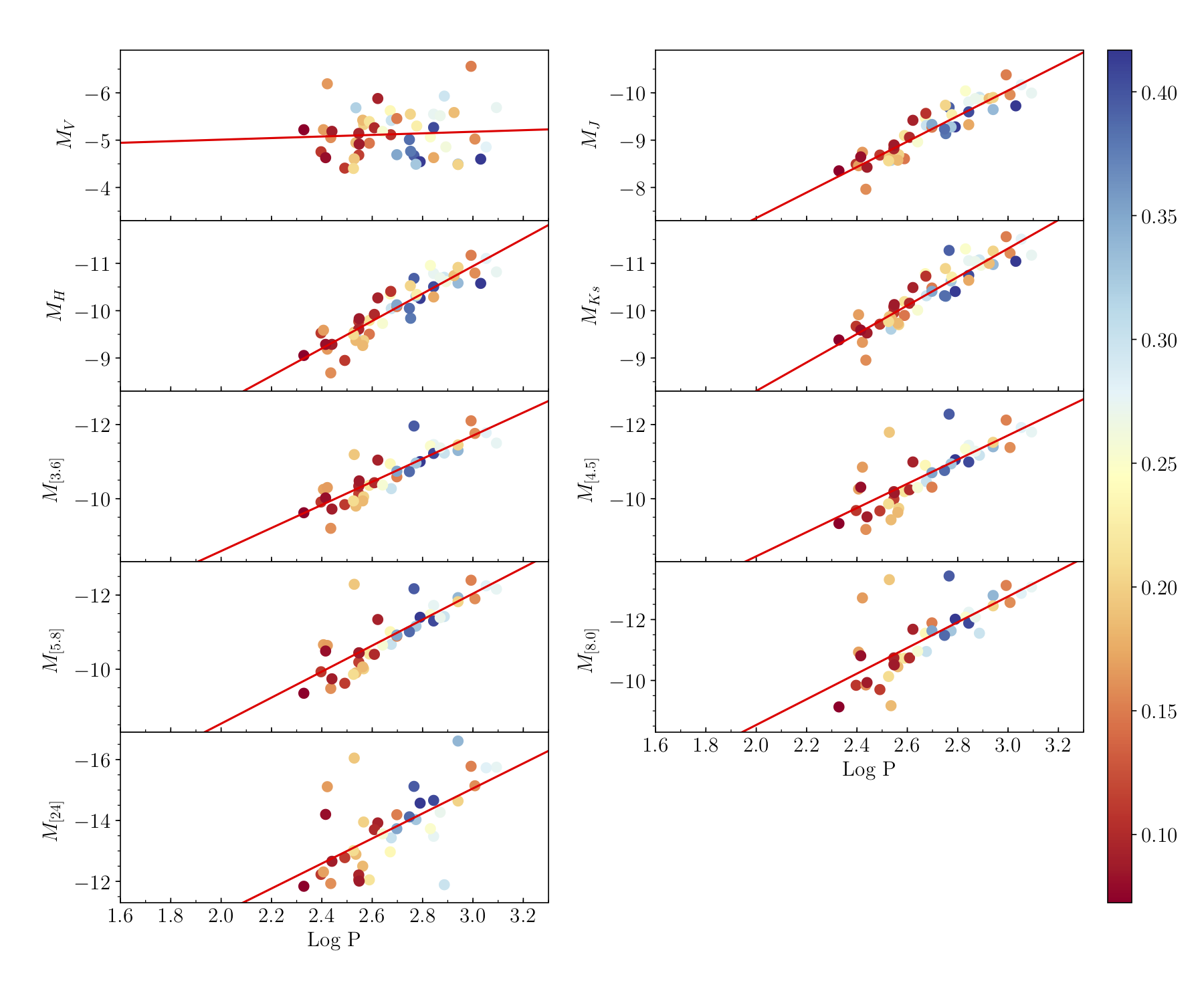}
	\caption{The same as Fig. \ref{fig:P_L_M33}, but for M31.}
	\label{fig:P_L_M31}
\end{figure}

\begin{figure}[ht]
	\centering
	\includegraphics[width=\textwidth]{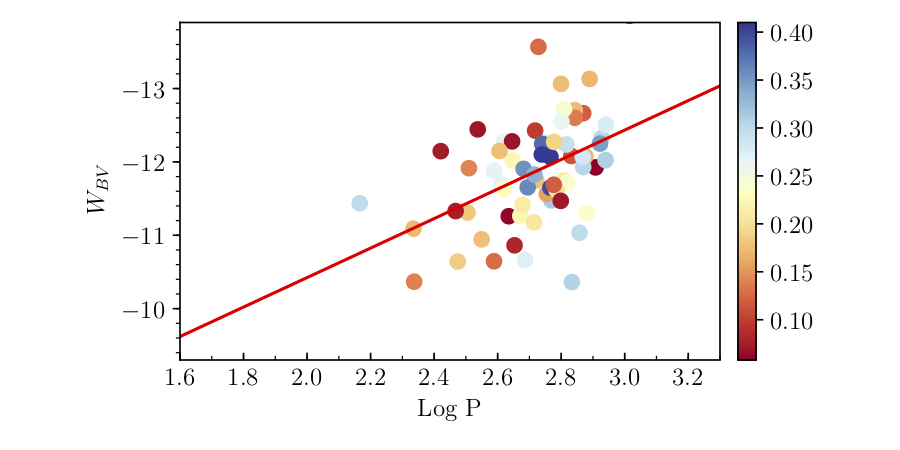}
	\caption{The Period-Luminosity relation in the $W_{BV}$ band for RSGs with semi-regular variation in M33. The symbols are the same as in Fig.\ref{fig:P_L_M33}}
	\label{fig:extinction_M33}
\end{figure}

\begin{figure}[ht]	
	\centering
	\includegraphics[width=\textwidth]{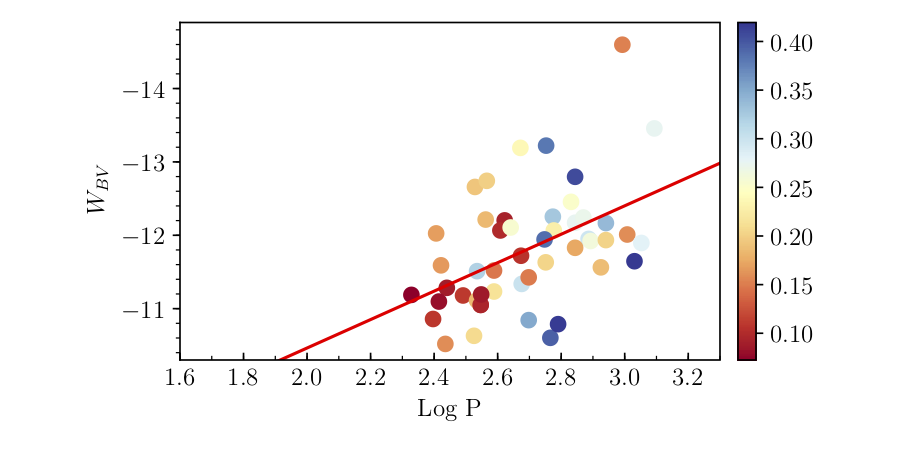}
	\caption{The same as Fig. \ref{fig:extinction_M33}, but for M31.}
	\label{fig:extinction_M31}
	
	\centering
	\includegraphics[width=\textwidth]{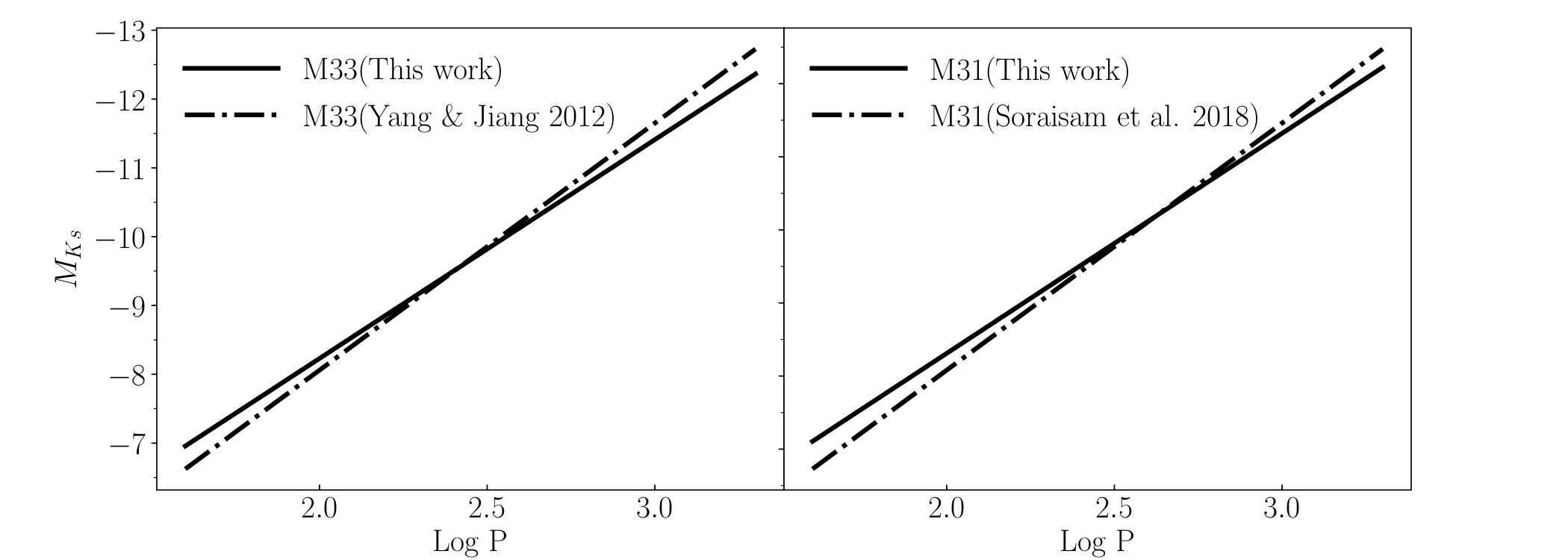}
	\caption{Comparison of the P-L relations of RSGs in M33 and M31 with previous works.}
	\label{fig:M31_M33_cmp}
\end{figure}

Following the convention, a linear function is taken to fit the relation between the absolute magnitude and $\log \rm P$.  The results are displayed in Fig.\ref{fig:P_L_M33}, Fig.\ref{fig:extinction_M33} and Table \ref{tab:M33_P_L} for the M33 sample, and in Fig.\ref{fig:P_L_M31}, Fig.\ref{fig:extinction_M31} and Table \ref{tab:M31_P_L} for the M31 sample. It is apparent that there is a tight relation in the near infrared and the mid-infrared band. Quantitatively the correlation coefficients are around 0.8 (see Table \ref{tab:M33_P_L} and Table \ref{tab:M31_P_L}), although it becomes less tight at the Spitzer [8.0] and [24] bands which is due to the scattering brought by circumstellar dust. On the other hand, the $V$ band shows almost no relation between brightness and period. The lack of P-L relation in the $V$ band may be caused by the large and inhomogeneous interstellar and circumstellar extinction, which is about 10 times that in the $K_{\rm S}$ band \citep{1989ApJ...345..245C, 2014ApJL..788...12W} in addition to the relatively large amplitude of variation. With no accurate measurement of interstellar extinction to individual star, we applied the Wesenheit index $W_{BV} = V - R_{V} (B - V)$. The parameter $R_{V}$ ($\equiv A_{V}/E(B - V)$, the total-to-selective extinction ratio) has an average value of 3.1 in the Milky Way, while it is not well determined in M31 and M33. This value is found to vary from about 2.0 to 3.3 towards different sightlines in M31 \citep{2014ApJ...785..136D,2015ApJ...815...14C}, and it is not well measured in M33. We take the value of 3.1 for simple and reasonably good approximation. Fig.\ref{fig:extinction_M33} and \ref{fig:extinction_M31} illustrate the relation of $W_{BV}$ with $\log \rm P$. The relation is much improved that the correlation coefficient is increased to around 0.4 (see the last row of Table \ref{tab:M33_P_L} and Table \ref{tab:M31_P_L}), which verifies that the lack of P-L relation in the $V$ band is partly caused by interstellar extinction. For a correct P-L relation in the $V$ band, an accurate correction of interstellar extinction should be made to each star individually and accurately. In summary, the P-L relation exists between the photosphere brightness and the period of light variation for RSGs with semi-regular variation in M33 and M31.  There is a tight P-L relation for apparent brightness in the 2MASS/$JHK_{\rm S}$ and Spitzer/IRAC1-3 bands. Meanwhile, the P-L relation should exist and can show up in the $V$ band when the interstellar extinction is corrected, and it will become tighter in the Spitzer [8.0] and [24] band when the emission of circumstellar dust is subtracted.


\section{Comparison with other results}
For further comparison with other results, we chose $K_{\rm S}$ band as a luminosity indicator and run Markov Chain Monte Carlo (MCMC) using PyMC3 \citep{2015arXiv150708050S} to obtain the uncertainty of the relation after taking the photometric error into account. The P-L relation in $K_{\rm S}$ band is as follows:

For M33, $M_{K_{\rm S}} = (-3.18 \pm 0.46) \times \rm log P + (-1.87 \pm 0.31)$.

For M31, $M_{K_{\rm S}} = (-3.01 \pm 0.38) \times \rm log P + (-2.29 \pm 0.24)$.

\subsection{M33}

\citet{2012ApJ...754...35Y} derived the P-L relation of 40 RSGs in M33 from a sample by \citet{1987AJ.....93..833K}: $M_{K} = (-3.59 \pm 0.41) \times \rm log P + (-0.88 \pm 0.69)$. The P-L relation is in agreement with this work within a reasonable error range as shown in Fig.\ref{fig:M31_M33_cmp}. However, we derived the P-L relation based on a larger sample of 84 sources with semi-regular variation and longer time baseline. Moreover, the P-L relation is derived in many more bands, including the visual $V$ band, other 2MASS bands and the Spitzer bands.

\subsection{M31}

The P-L relation of RSGs in M31 was obtained previously in the $K$ band by \citet{2018ApJ...859...73S} who also used the iPTF data. Their data spanned from MJD 56000 to 58000. Here we used the iPTF data from MJD 55050 to 57050 by the limit to data access.
\citet{2018ApJ...859...73S} compared with the theoretical P-L relation from Modules for Experiments in Stellar Astrophysics (MESA) \citep{2011ApJS...192..3P,2013ApJS...208..4P,2015ApJS...220..15P,2018ApJS...234..34P}. Consistent with our results, the majority of sources pulsate in the fundamental mode. After ignoring 10 RSGs in the first overtone pulsation mode and 2 RSGs with periods $<$ 100 days, they obtained the P-L relation: $M_{K} = (-3.38 \pm 0.27) \times \rm log P + (-1.32 \pm 0.75)$. This P-L relation is also in agreement with our work within the error range as shown in Fig.\ref{fig:M31_M33_cmp}.

\subsection{The Galaxy, LMC, SMC}

\begin{figure}[ht]		
	\centering
	\includegraphics[width=\textwidth]{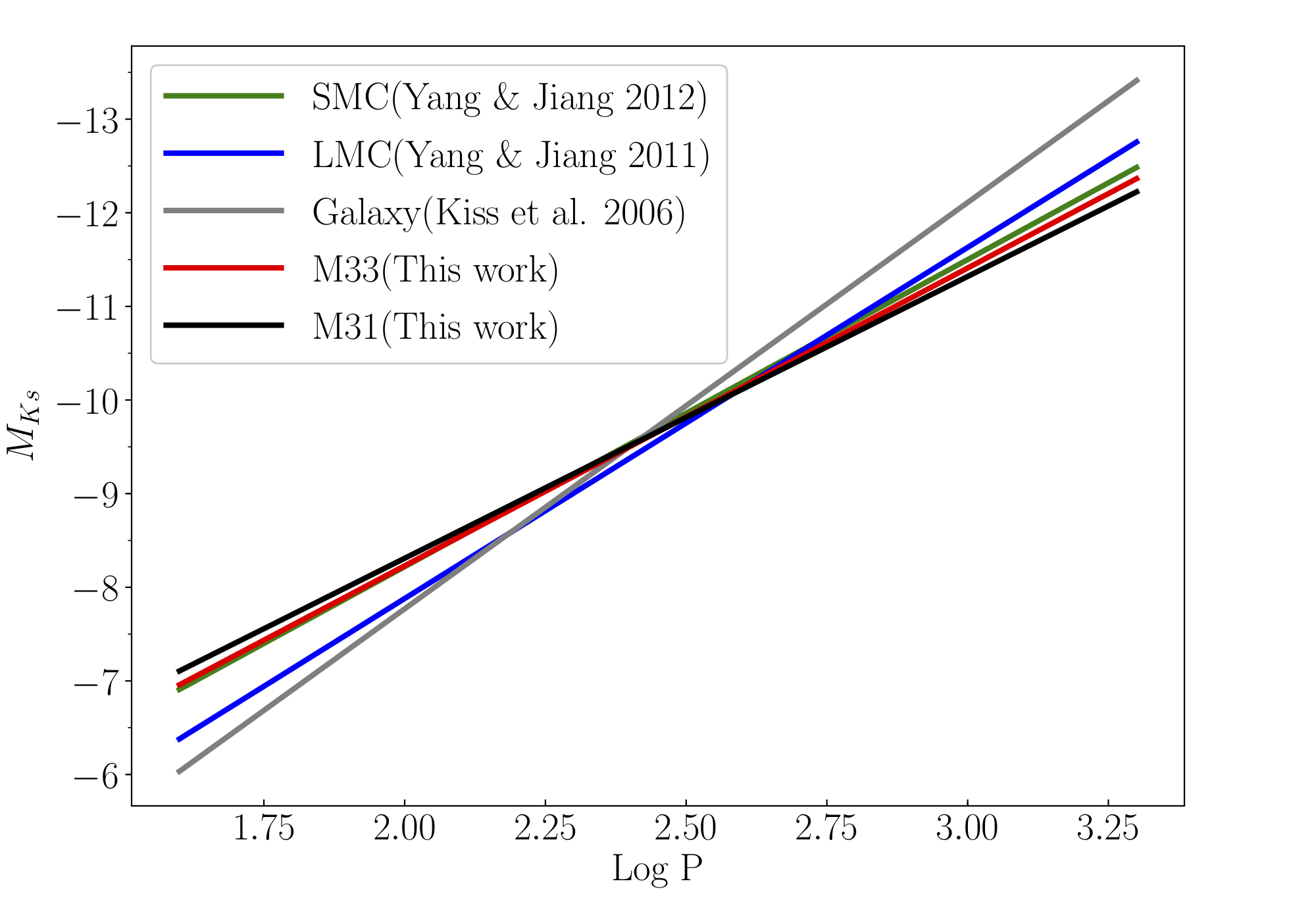}
	\caption{The P-L relations of RSGs in nearby galaxies and the Milky Way.}
	\label{fig:result_cmp}
\end{figure}

\citet{2006MNRAS.372.1721K} derived P-L relation based on 18 RSGs in the Milky Way and the light curves collected by AAVSO. Most of these RSGs are members of associations and clusters, and their distances are taken as that of the host associations and clusters. For three sources they used the Hipparcos parallaxes.
Based mainly on the ASAS and MACHO data, \citet{2011ApJ...727...53Y,2012ApJ...754...35Y} derived the P-L relation of 47 RSGs in LMC and 21 RSGs in SMC.
Comparison of the P-L relations of RSGs in different galaxies are shown in Fig.\ref{fig:result_cmp}. It can be seen that the P-L relation in M33 is very similar to SMC. The P-L relations seem to be similar in these galaxies except for the Milky Way. The P-L relation of RSGs in the Milky Way appears a little different that may be caused by the dispersion of estimated distance and interstellar extinction.
It may be argued that the P-L relations in K band is universal in these nearby galaxies. It should be mentioned that the optical bands would be better to examine the dependence of the P-L relation on metallicity because the P-L relation is less dependent on metallicity in near-infrared bands than optical bands \citep{2017MNRAS.466.2805B}. Nevertheless, interstellar extinction is serious  to make the determination of P-L relation more uncertain in optical bands.

\begin{figure}[ht]		
	\centering
	\includegraphics[width=\textwidth]{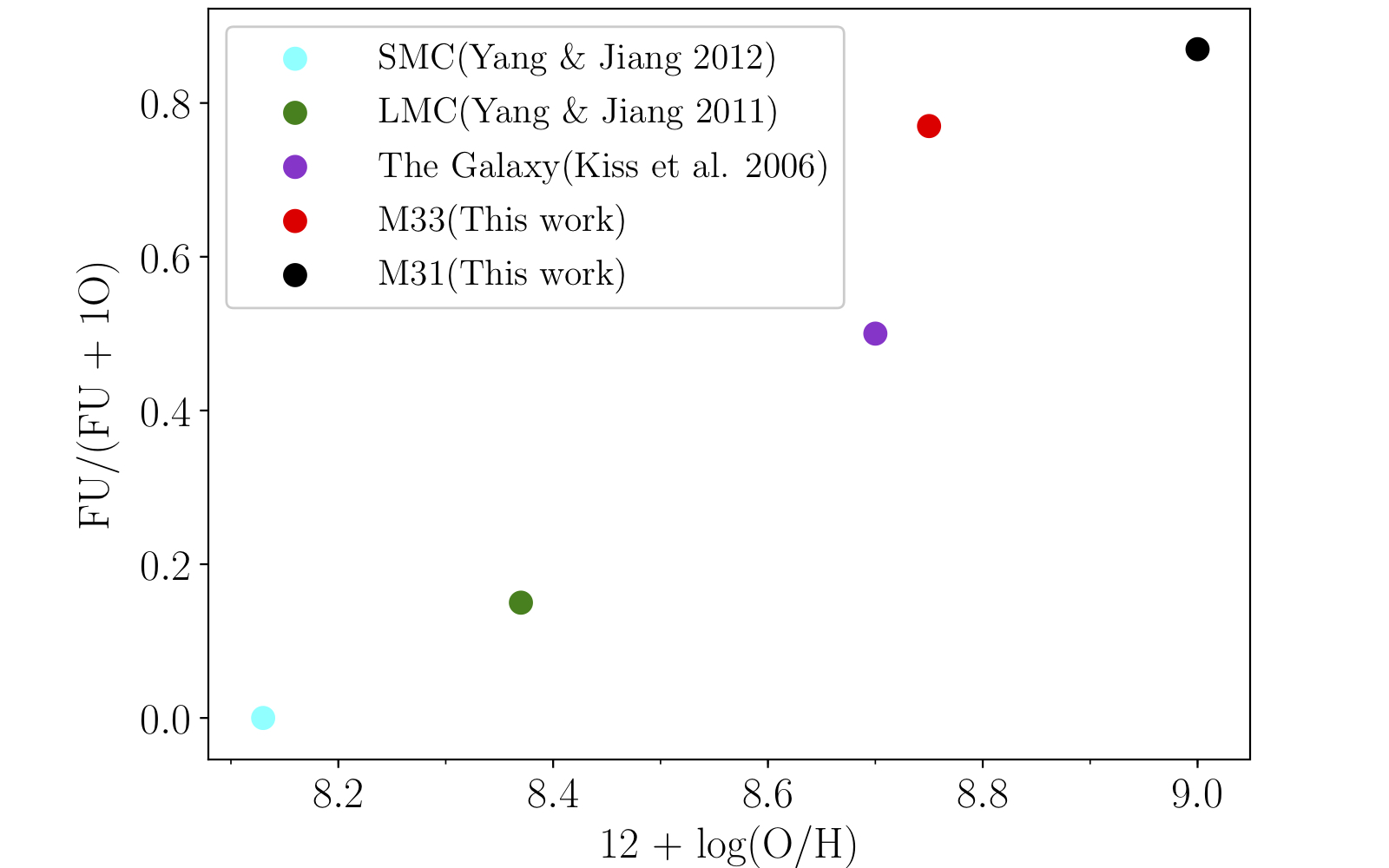}
	\caption{The number ratio of FU/(FU + 1O) versus metallicity in nearby galaxies and the Milky Way Galaxy.}
	\label{fig:FU_1O}
\end{figure}

The ratio of $\rm N_{FU}$ (the number of RSGs which pulsate in FU pulsation mode) to $\rm N_{1O}$ (the number of RSGs which pulsate in 1O pulsation mode) increase with the metallicity. The values $\rm N_{FU}/(N_{FU} + N_{1O})$ vary from SMC (0\%), LMC (15\%), the Galaxy (50\%), M33 (77\%) to M31 (87\%) as shown in Fig.\ref{fig:FU_1O} when we compare the results of SMC, LMC, Galaxy from \citet{2012ApJ...754...35Y}. The percentages may suffer some uncertainty because the identification of pulsating mode depends on the theoretical models adopted and the borderline between the two modes are not very sharp. However, it is apparent that $\rm N_{FU}/(N_{FU} + N_{1O})$  increases with metallicity. We explored the underlying physics and found that this may be related to convection. \citet{1998A&A...336..553Y} calculated the turbulent convective Cepheid models and found that the 1O mode would be suppressed at a large mixing length in comparison with the FU mode, which may be understood because the 1O mode occurs in relatively shallower layer and would be influenced more by the envelop convection. Although this model is for Cepheids, it should be applicable to RSGs since RSGs have even stronger convection due to lower effective temperature. Furthermore, \citet{2018ApJ...853...79C} found that the mixing length increases with metallicity for RSGs. This may be caused by that the high metallicity leads to less effective radiative transfer due to higher opacity, and consequently stronger convection is needed. Combining the two theories --- the 1O mode would be suppressed at large mixing length and the mixing length increases with metallicity, we can expect that the 1O mode pulsation occurs less frequently at high metallicity. Specifically, the 1O mode pulsators should be the least frequent and thus largest FU/(FU+1O) in M31 with the highest metallicity.

\section{Summary and Conclusion}

The study is based on a relatively complete and pure sample of RSGs in M33 and M31. From the preliminary samples of RSGs (749 in M33 and 437 in M31) which were selected by the visual colors with the LGGS multiband photometry, foreground dwarf and giant stars (2 in M33 and 5 in M31) are further excluded according to their distance modulus from the GAIA/DR2 catalog. The pollution of AGB stars in the local galaxy is carefully checked by the locations in several CMDs, which found almost no contamination by AGB stars while the dust emission has the significant influence on the flux in the mid-infrared bands such as the Spitzer/[8.0] and [24] bands. Consequently, we find that there are 717 RSGs in M33 and 420 RSGs in M31 which form the basis to analyze the variation of RSGs.


The variation is analyzed with the time-series data taken by the iPTF survey, spanning about 2000 days. From the CCD images, the differential photometry is performed to all the RSGs with the Sextractor tool by making the average of all RSGs as the reference star, which achieves an accuracy of about 0.04 mag. The light curve is then fitted by the PDM, GLS, DFT and WWZ methods, and a period is considered to be real only when it is identified by at least two methods for RSGs with semi-regular variation. Based on the characteristics of lightcurves, RSGs are classified into semi-regular variables, variables with LSP, irregular variables and others. In total, 84 RSGs in M33 and 56 RSGs in M31 are found to show semi-regular variation, in which, 19 stars in M33 and 7 stars in M31 are identified in first overtone mode, and excluded for P-L analysis. 65 RSGs in M33 and 49 RSGs in M31 are taken to analyze the P-L relation in visual, near-infrared and mid-infrared bands.

The P-L relation exists in all the infrared bands. In particular, the relation is very tight in the 2MASS/$JHK_{\rm S}$ bands and the Spitzer/IRAC1-3 bands.
In the Spitzer/IRAC [8.0] and Spitzer/MIPS [24] bands, the P-L relation is relatively sparse due to the emission of circumstellar dust. In the visual band, the P-L relation is not clear which can be understood by inhomogeneous extinction. When the de-reddening Wesenheit index $W_{BV} = V - R_{V} (B-V)$ is taken into account, the P-L relation shows up although it is not as tight as in the infrared bands.

The results are compared with the P-L relations of RSGs in other galaxies, i.e. SMC, LMC and the Milky Way. It is found that the P-L relation in K band in these galaxies are similar, showing no apparent dependence on the metallicity.

From the P-L relation, the pulsation mode of RSGs in the M33 and M31 is compared with the theoretical model and found to be very possibly in the fundamental pulsation mode. When compared with the SMC, LMC and Galaxy, it seems that the pulsation mode shows dependence on the metallicity. This phenomenon may be related to convection and its dependence on metallicity and deserves further investigation.

Due to the limitations of the iPTF data available at now, the LSP of RSGs cannot be determined with reasonable accuracy. We may obtain long baseline data from the combination of the iPTF data and the Panoramic Survey Telescope and Rapid Response System (Pan-STARRS) DR2 data to determine the LSP in future.

\begin{acknowledgements}
We thank Prof. Yan Li for helpful discussions. Special thanks go to the anonymous referee for his/her very helpful suggestions, which improved the paper significantly.
This work is supported by NSFC through Projects 11533002 and U1631104.
This work made use of the data taken by \emph{LGGS}, iPTF, 2MASS and Spitzer.
This work made use of PyAstronomy.
\end{acknowledgements}

%

\vspace{5mm}


\software{astropy \citep{2013A&A...558A..33A},
          	SExtractor \citep{1996A&AS..117..393B},
            VARTOOLS \citep{2016A&C....17....1H}
          }



\begin{figure}[ht]	
	\centering
	\includegraphics[width=\textwidth]{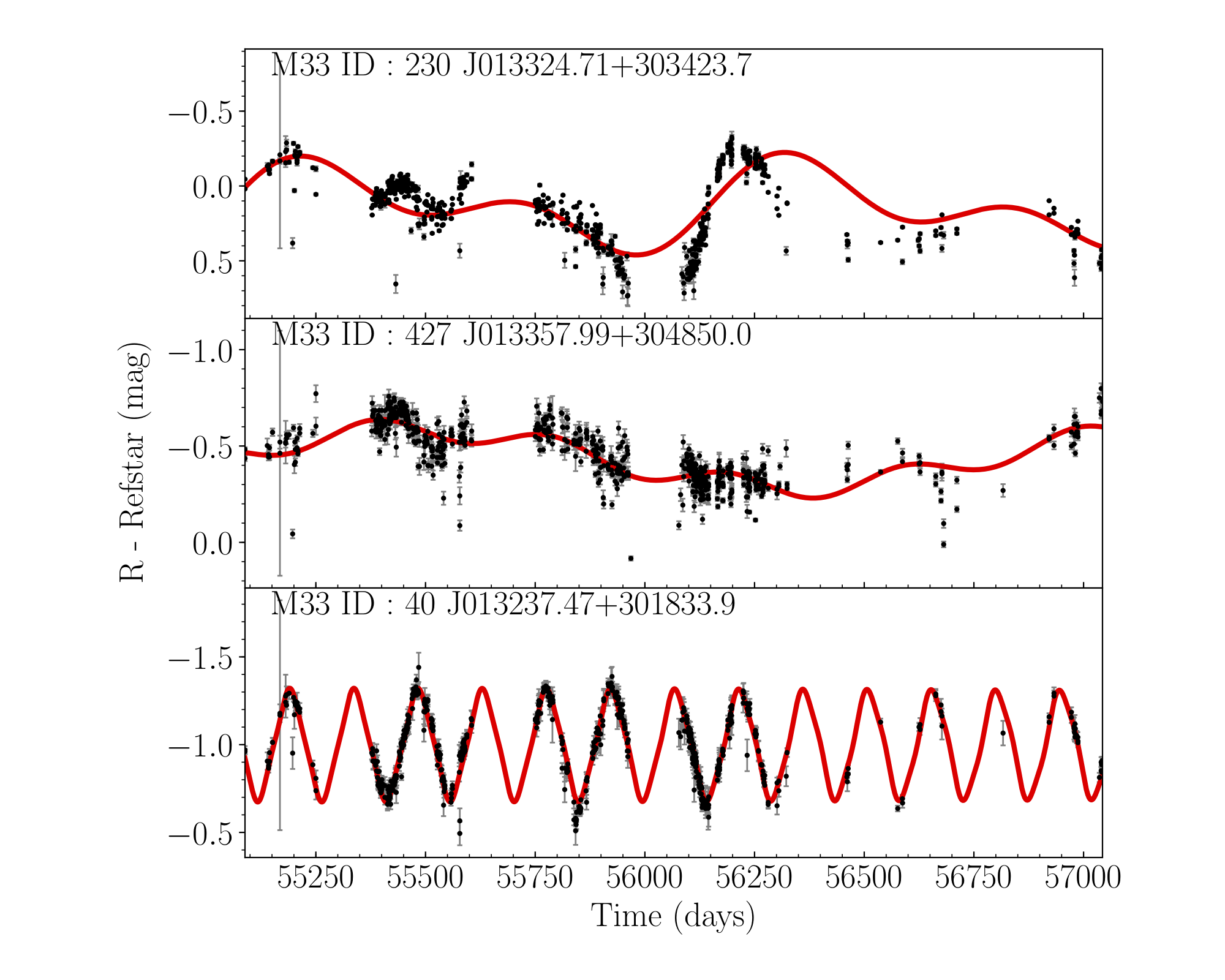}
	\caption{Three categories of lightcurve. From top to bottom are the irregular, long secondary period and semi-regular variables.}
	\label{fig:type}
\end{figure}

\FloatBarrier

\begin{deluxetable*}{cccccc}[ht]
	\tablecaption{Linear fit results between the bolometric magnitude and the multiband magnitude from RSG candidates in M33. The fitting form follows by $m(\lambda) = a * m_{\rm bol}(\lambda) + b$ \label{tab:M33_mutilbandfit_tab}}.

	\tablewidth{0pt}
	\tablehead{
		\colhead{Band} & \colhead{Slope} & \colhead{Intercept} & \colhead{Correlation} &
		\colhead{$m_{\rm max}$} & \colhead{$m_{\rm min}$}\\
	}
	\startdata
	$J$ & 1.09 	&-2.90 	&0.95 	&13.54 	&18.73  \\
	$H$ & 1.09 	&-3.71 	&0.94 	&12.73 	&17.93  \\
	$K_{\rm S}$ & 1.15 	&-5.13 	&0.97 	&12.28 	&17.79  \\
	\begin{math}[3.6]\end{math} & 1.25 	&-6.80 	&0.86 	&12.02 	&17.97  \\
	\begin{math}[4.5]\end{math} & 1.32 	&-7.87 	&0.85 	&12.06 	&18.36  \\
	\begin{math}[5.8]\end{math} & 1.41 	&-9.75 	&0.79 	&11.46 	&18.17  \\
	\begin{math}[8.0]\end{math} & 1.34 	&-9.57 	&0.58 	&10.61 	&16.99  \\
	\begin{math}[24]\end{math} & 1.09 	&-7.55 	&0.45 	&8.84 	&14.02  \\
	\enddata
\end{deluxetable*}

\begin{deluxetable*}{cccccc}[ht]
	\tablecaption{Linear fit results between the bolometric magnitude and the multiband magnitude from RSG candidates in M31. The fitting form follows by $m(\lambda) = a * m_{\rm bol}(\lambda) + b$ \label{tab:M31_mutilbandfit}}.
	\tablewidth{0pt}
	\tablehead{
		\colhead{Band} & \colhead{Slope} & \colhead{Intercept} & \colhead{Correlation} &
		\colhead{$m_{\rm max}$} & \colhead{$m_{\rm min}$}\\
	}
	\startdata
	$J$ & 1.08 & -2.85 & 0.95 & 13.19 & 18.35\\
	$H$ & 1.11 & -4.08 & 0.95 & 12.33 & 17.61\\
	$K_{\rm S}$ & 1.17 & -5.43 & 0.98 & 11.88 & 17.45\\
	\begin{math}[3.6]\end{math} & 1.19 & -6.10 & 0.88 & 11.58 & 17.27\\
	\begin{math}[4.5]\end{math} & 1.32 & -8.01 & 0.85 & 11.51 & 17.79\\
	\begin{math}[5.8]\end{math} & 1.30 & -7.99 & 0.85 & 11.29 & 17.49\\
	\begin{math}[8.0]\end{math} & 1.48 & -11.41 & 0.78 & 10.58 & 17.66\\
	\begin{math}[24]\end{math} & 1.01 & -5.73 & 0.53 & 9.22 & 14.03\\
	\enddata
\end{deluxetable*}

\begin{deluxetable*}{ccccc}[ht]
	\tablecaption{The parameters used for the period determination \label{tab:The parameters used for the period determination}}
	\tablewidth{0pt}
	\tablehead{
	\colhead{Input parameters} & \colhead{GLS} & \colhead{DFT} & \colhead{WWZ} & \colhead{PDM} \\
	}
	\startdata
	Minimum period & 10 & 10 & 10 & 10 \\
	Maximum period &3500 & 3500 & -- & 3500 \\
	Period step & 0.01 & 0.01 & -- & 0.50 \\
	Time-Shift step & -- & -- & 20 & -- \\
	Frequency sampling & -- & -- & 0.5714 & -- \\
	Software & VARTOOLS & VARTOOLS & VARTOOLS & PyAstronomy
	\enddata
\end{deluxetable*}

\begin{deluxetable*}{cccccc}[ht]
	\tablecaption{The parameters of the P-L relations of RSGs in M33 \label{tab:M33_P_L}}
	\tablewidth{0pt}
	\tablehead{
		\colhead{Band} & \colhead{Slope} & \colhead{Intercept} & \colhead{Correlation} \\
	}
	\startdata
	$M_{V}$ & -0.23 & -4.99 & -0.07\\
	$M_{J}$ & -2.69 & -2.08 & -0.79\\
	$M_{H}$ & -3.02 & -1.99 & -0.84\\
	$M_{K_{\rm S}}$ & -3.18 & -1.87 & -0.85\\
	$M_{[3.6]}$ & -3.84 & -0.24 & -0.85\\
	$M_{[4.5]}$ & -4.15 & 0.73 & -0.82\\
	$M_{[5.8]}$ & -4.57 & 1.42 & -0.82\\
	$M_{[8.0]}$ & -5.01 & 1.74 & -0.75\\
	$M_{[24]}$ & -6.32 & 3.31 & -0.73\\
	$W_{BV}$ & -2.01 & -6.40 & -0.46\\
	\enddata
	\tablecomments{This table shows the parameters slope, intercept and correlation from Period-Luminosity relation of RSGs in M33.}
\end{deluxetable*}

\begin{deluxetable*}{cccccc}[ht]
	\tablecaption{The parameters of the P-L relations of RSGs in M31 \label{tab:M31_P_L}}
	\tablewidth{0pt}
	\tablehead{
		\colhead{Band} & \colhead{Slope} & \colhead{Intercept} & \colhead{Correlation} \\
	}
	\startdata
	$M_{V}$ & -0.17 & -4.68 & -0.07\\
	$M_{J}$ & -2.69 & -1.97 & -0.92\\
	$M_{H}$ & -2.90 & -2.25 & -0.91\\
	$M_{K_{\rm S}}$ & -3.01 & -2.29 & -0.92\\
	$M_{[3.6]}$ & -3.11 & -2.36 & -0.87\\
	$M_{[4.5]}$ & -3.26 & -1.92 & -0.80\\
	$M_{[5.8]}$ & -3.51 & -1.52 & -0.83\\
	$M_{[8.0]}$ & -4.22 & -0.10 & -0.74\\
	$M_{[24]}$ & -4.11 & -2.73 & -0.63\\
	$W_{BV}$ & -1.94 & -6.59 & -0.48\\
	\enddata
	\tablecomments{This table shows the parameters slope,intercept and correlation from Period-Luminosity relation of RSGs in M31.}
\end{deluxetable*}
\FloatBarrier

\appendix

\section{Lightcurve of RSGs with Semi-Regular Variation in M33 and M31} \label{appendix A}

\begin{figure}[ht]
	\centering
	\includegraphics[width=\textwidth]{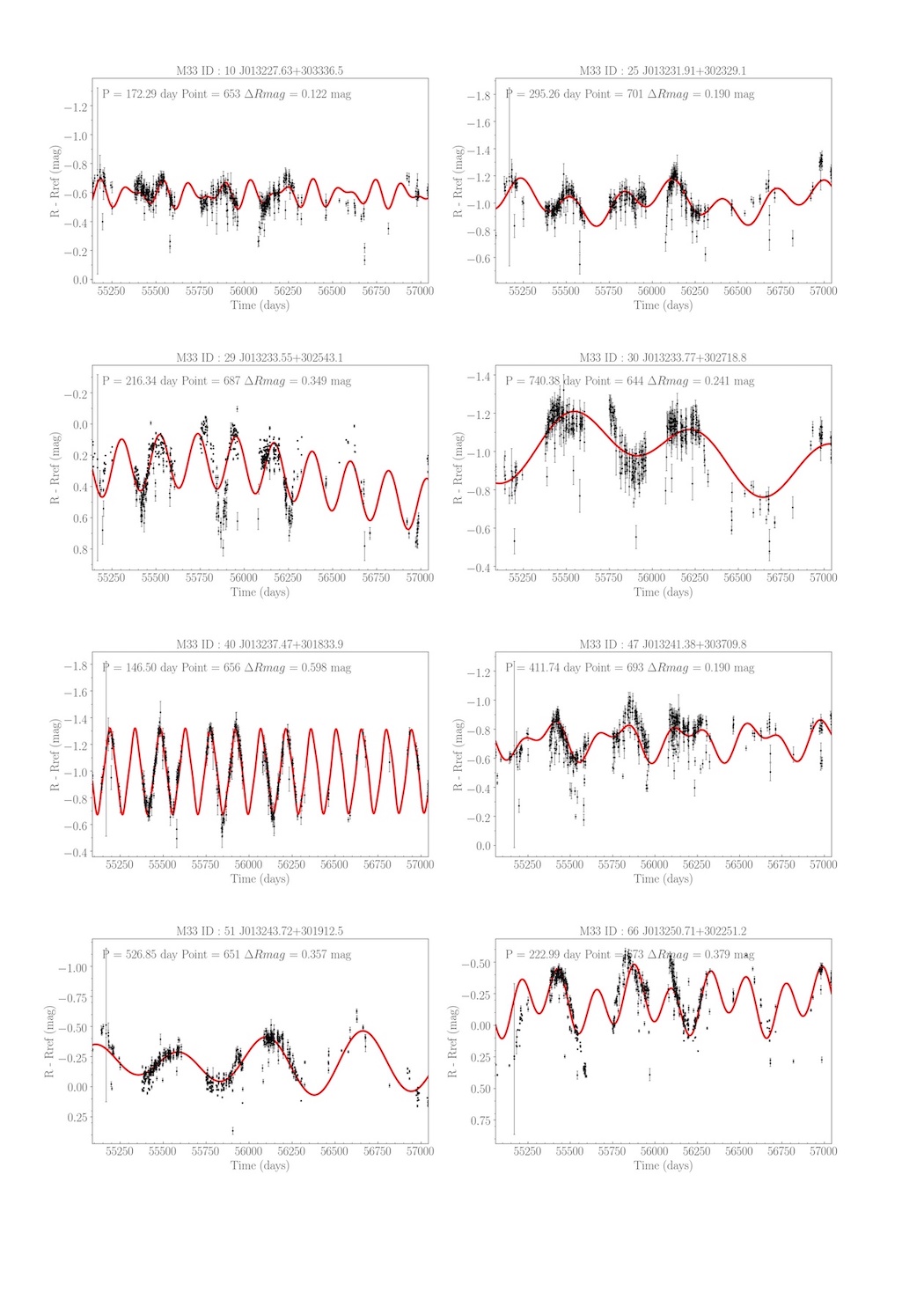}
\end{figure}

\begin{figure}[ht]
	\centering
	\includegraphics[width=\textwidth]{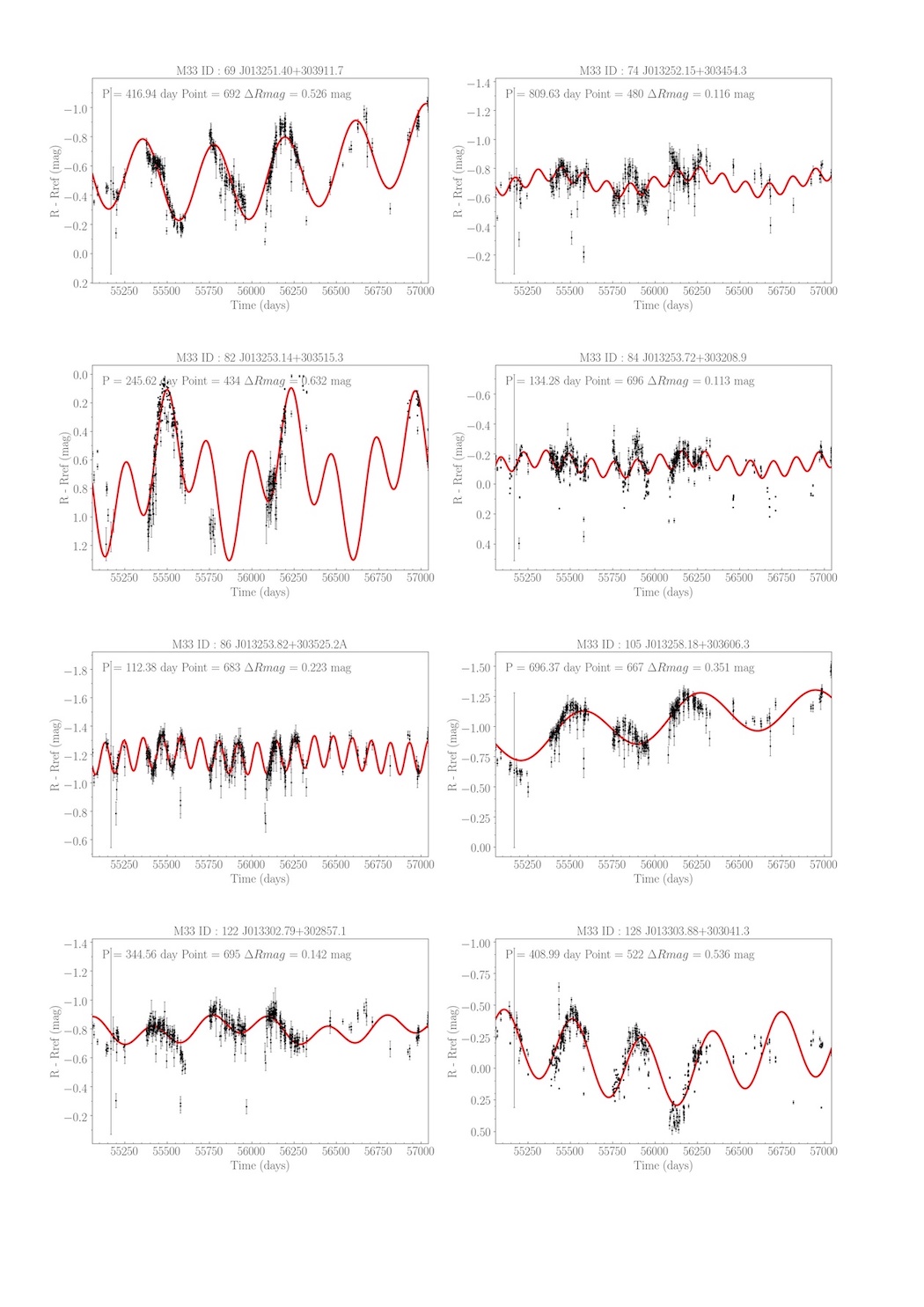}
\end{figure}

\begin{figure}[ht]
	\centering
	\includegraphics[width=\textwidth]{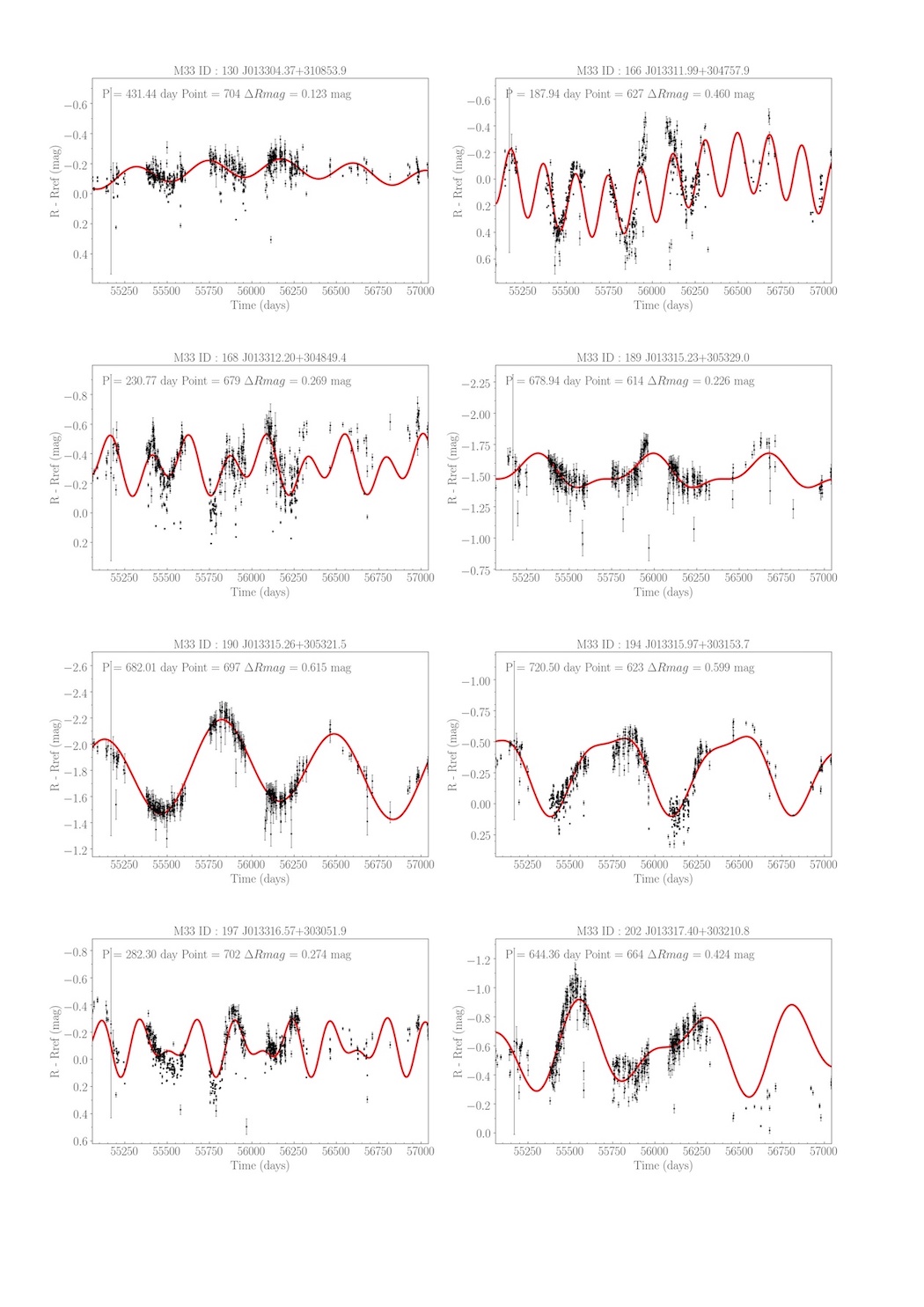}
\end{figure}

\begin{figure}[ht]
	\centering
	\includegraphics[width=\textwidth]{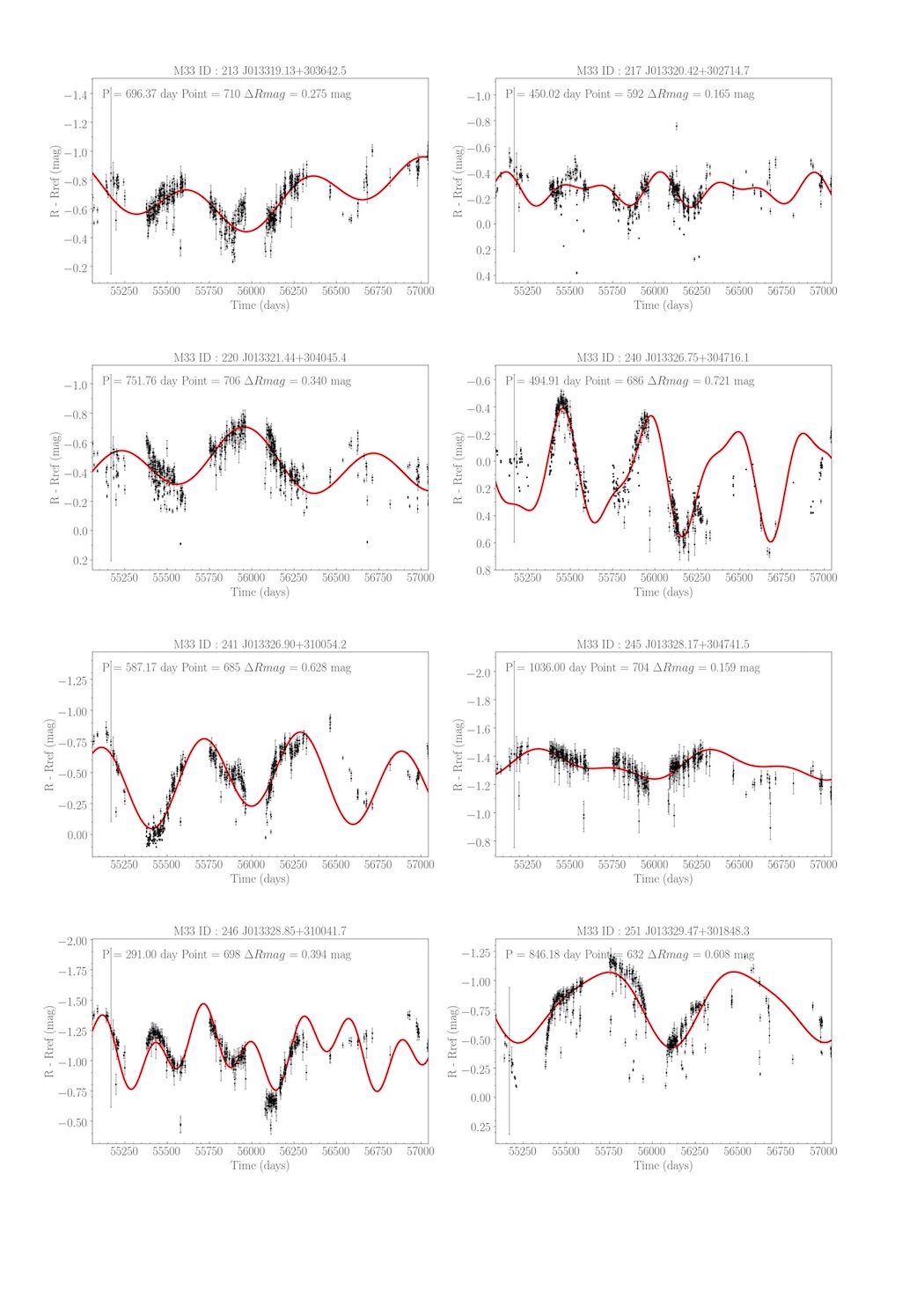}
\end{figure}

\begin{figure}[ht]
	\centering
	\includegraphics[width=\textwidth]{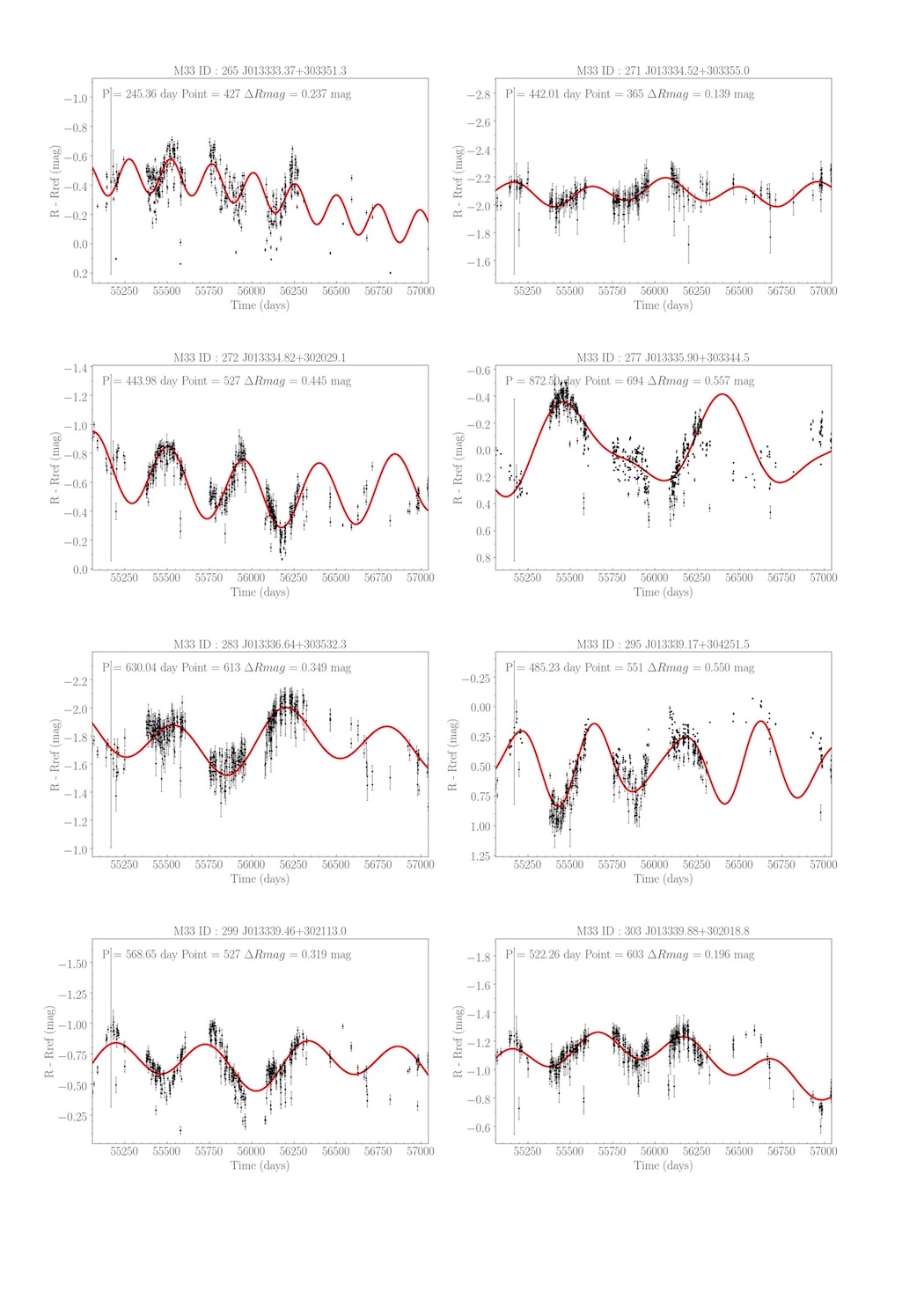}
\end{figure}

\begin{figure}[ht]
	\centering
	\includegraphics[width=\textwidth]{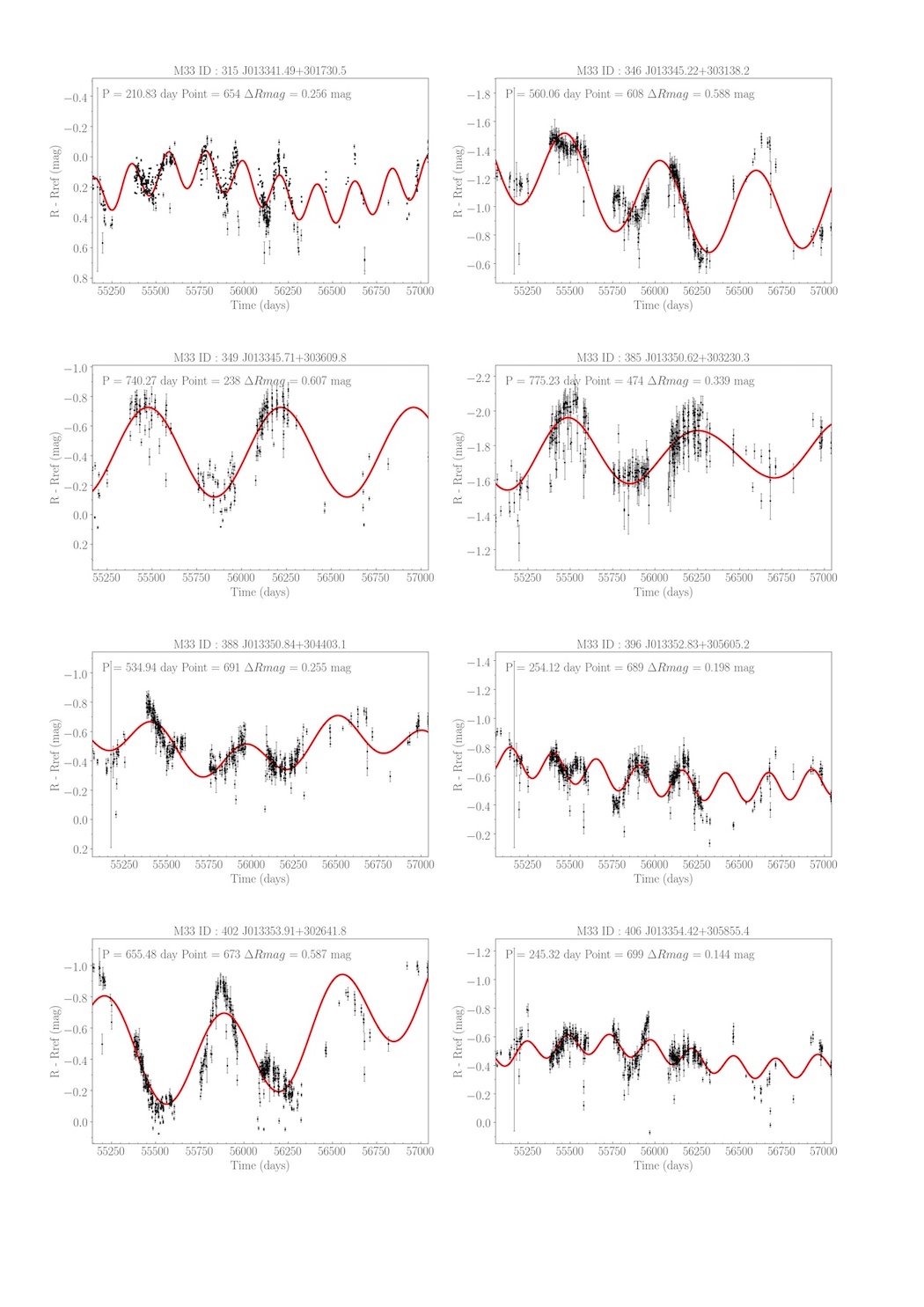}
\end{figure}

\begin{figure}[ht]
	\centering
	\includegraphics[width=\textwidth]{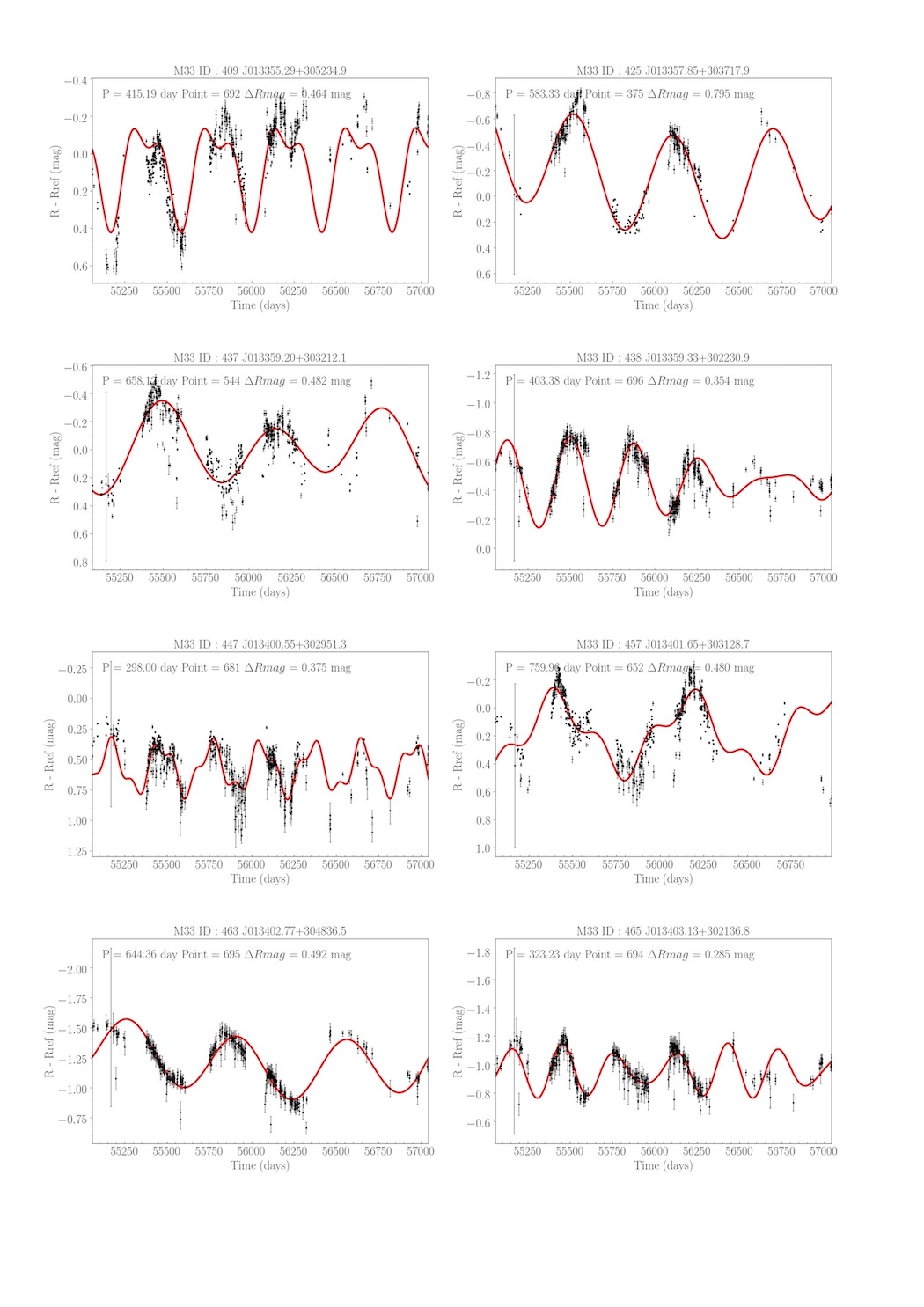}
\end{figure}

\begin{figure}[ht]
	\centering
	\includegraphics[width=\textwidth]{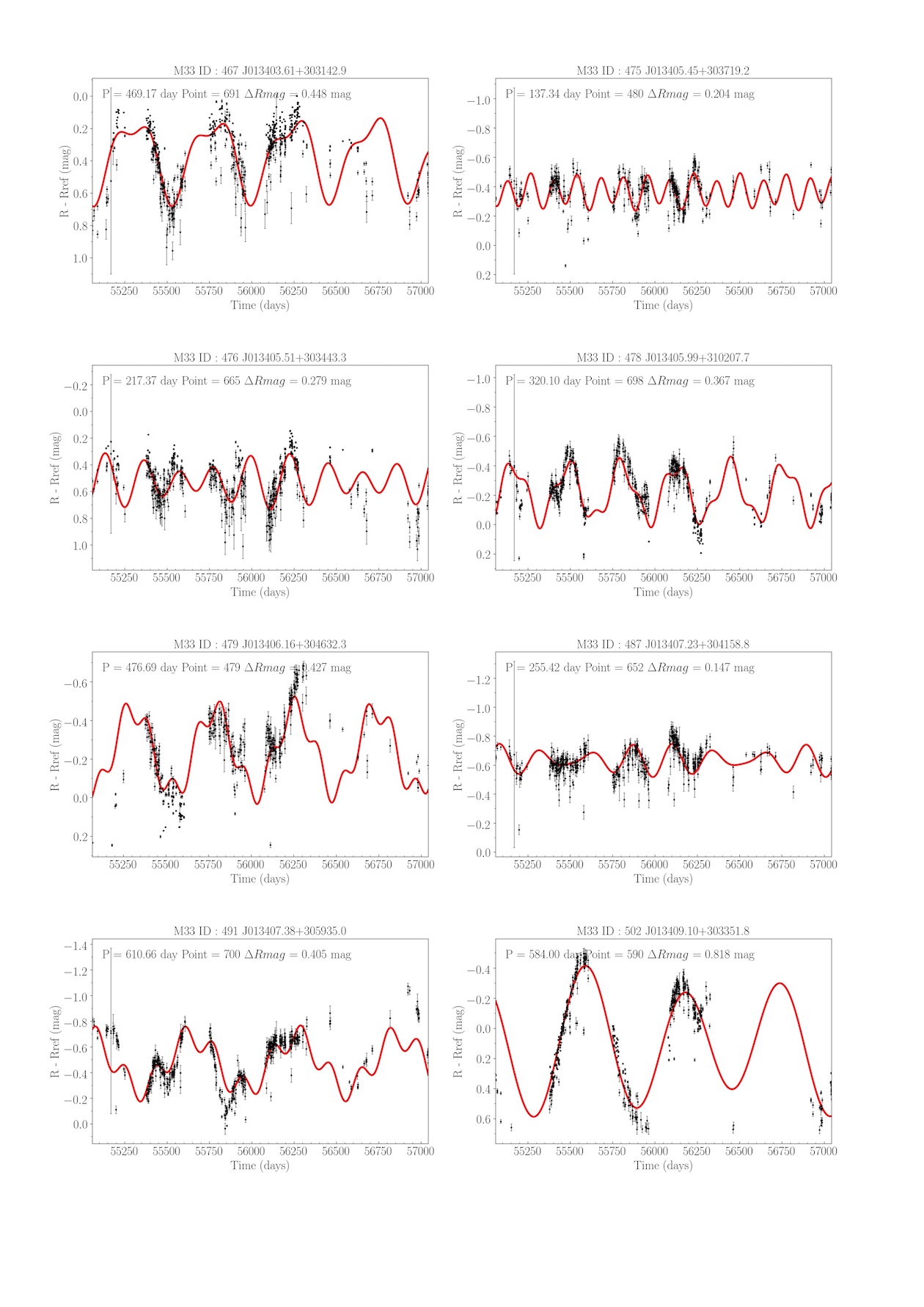}
\end{figure}

\begin{figure}[ht]
	\centering
	\includegraphics[width=\textwidth]{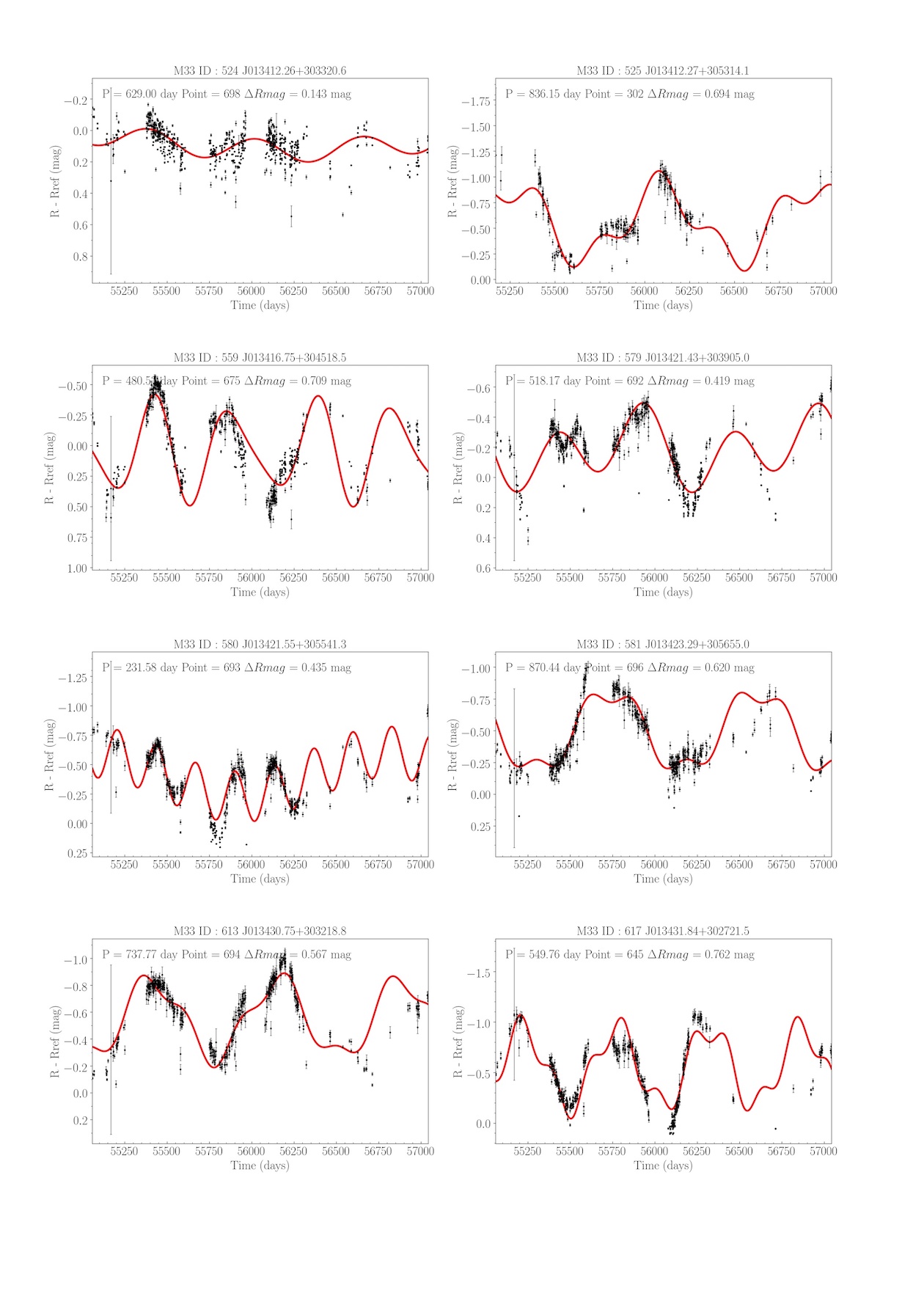}
\end{figure}

\begin{figure}[ht]
	\centering
	\includegraphics[width=\textwidth]{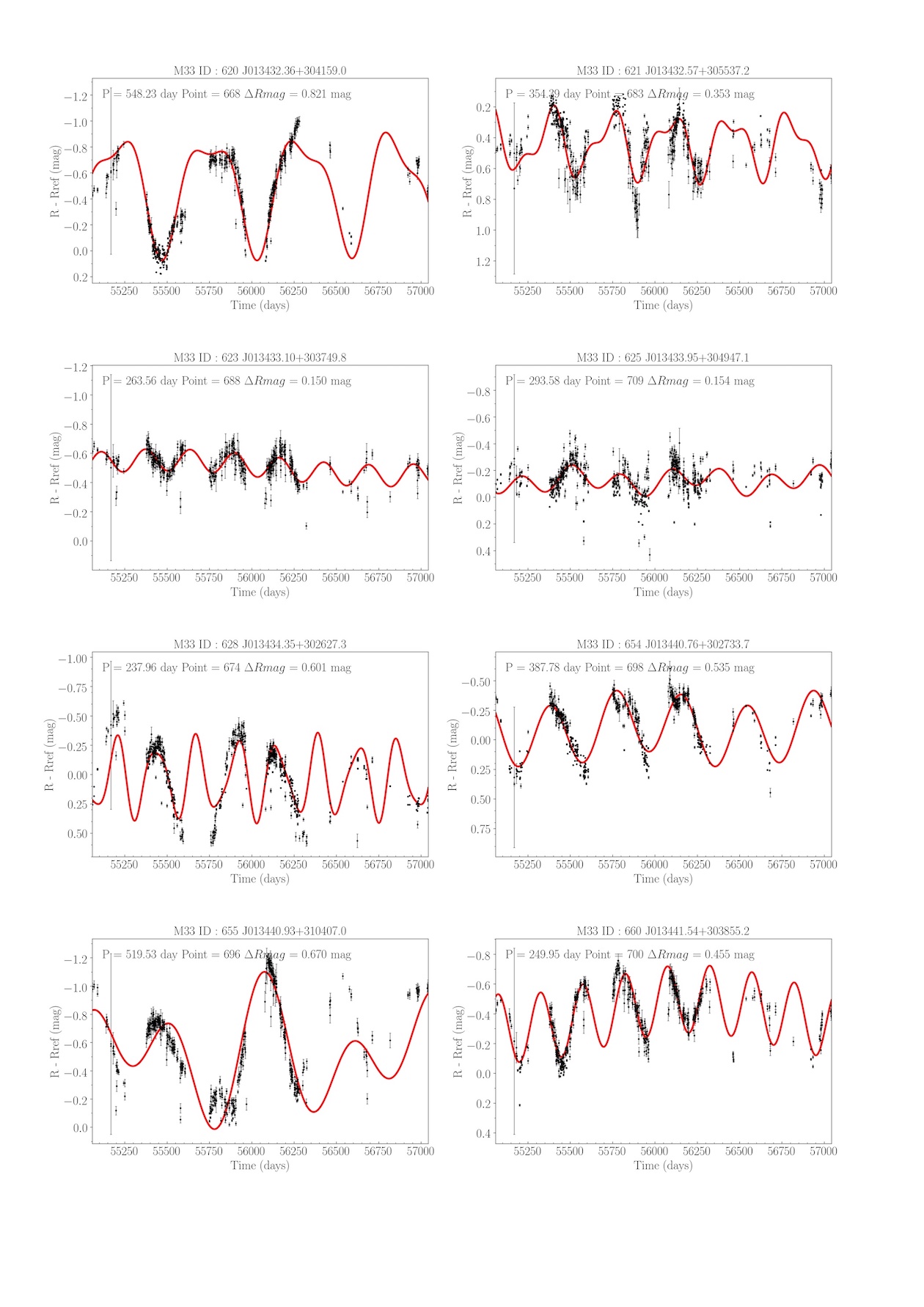}
\end{figure}

\begin{figure}[ht]
	\centering
	\includegraphics[width=\textwidth]{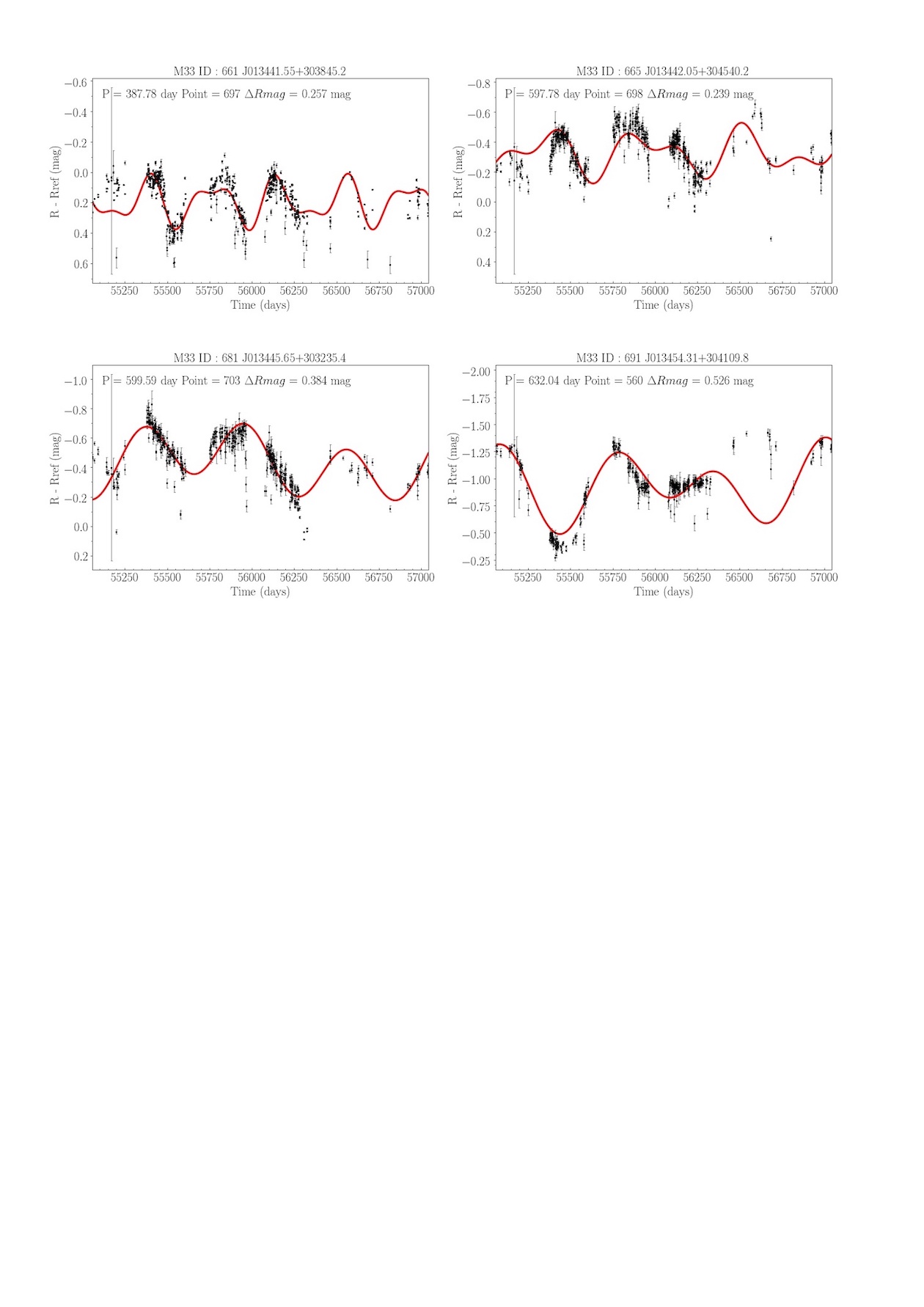}
\end{figure}

\begin{figure}[ht]
	\centering
	\includegraphics[width=\textwidth]{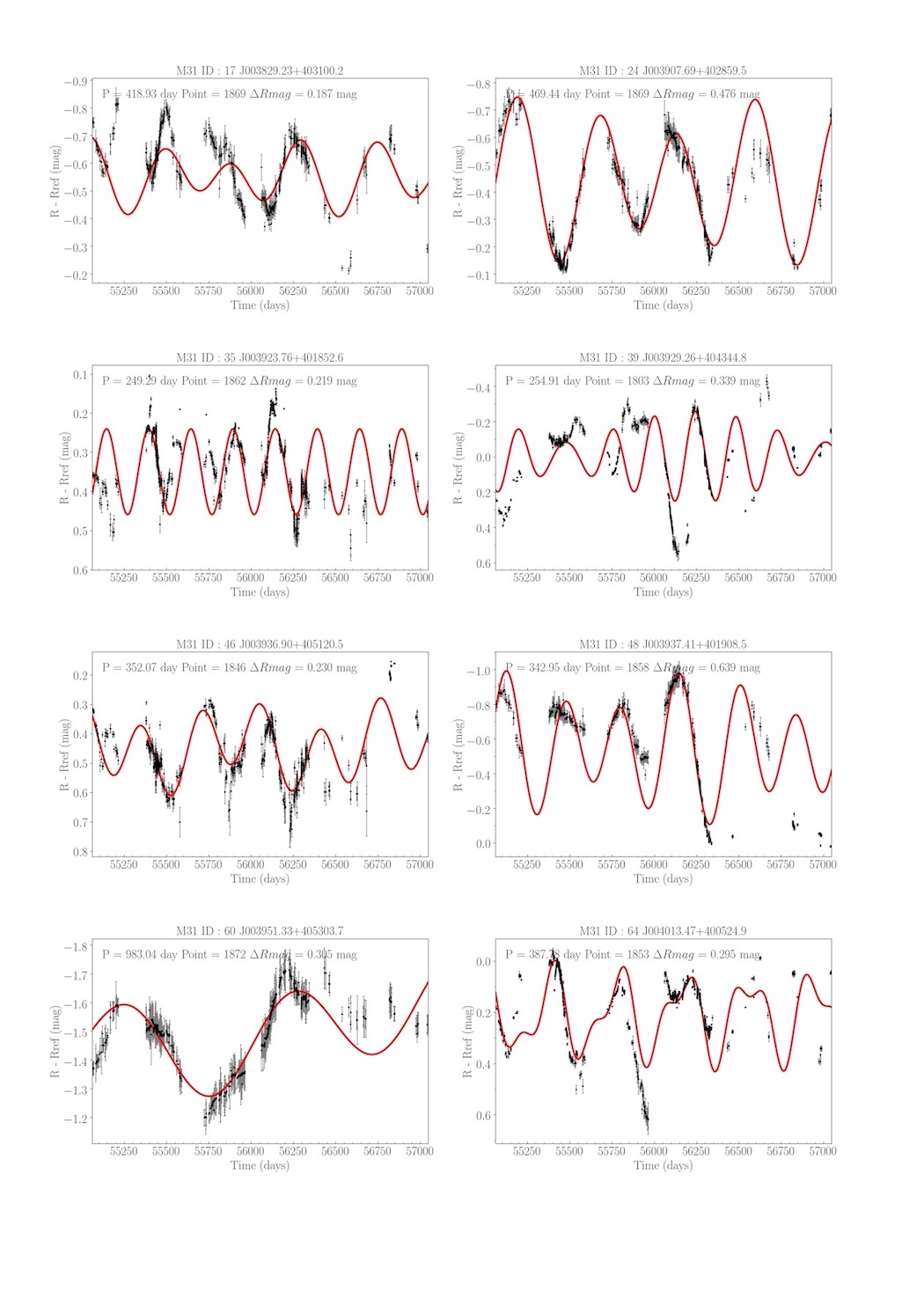}
\end{figure}

\begin{figure}[ht]
	\centering
	\includegraphics[width=\textwidth]{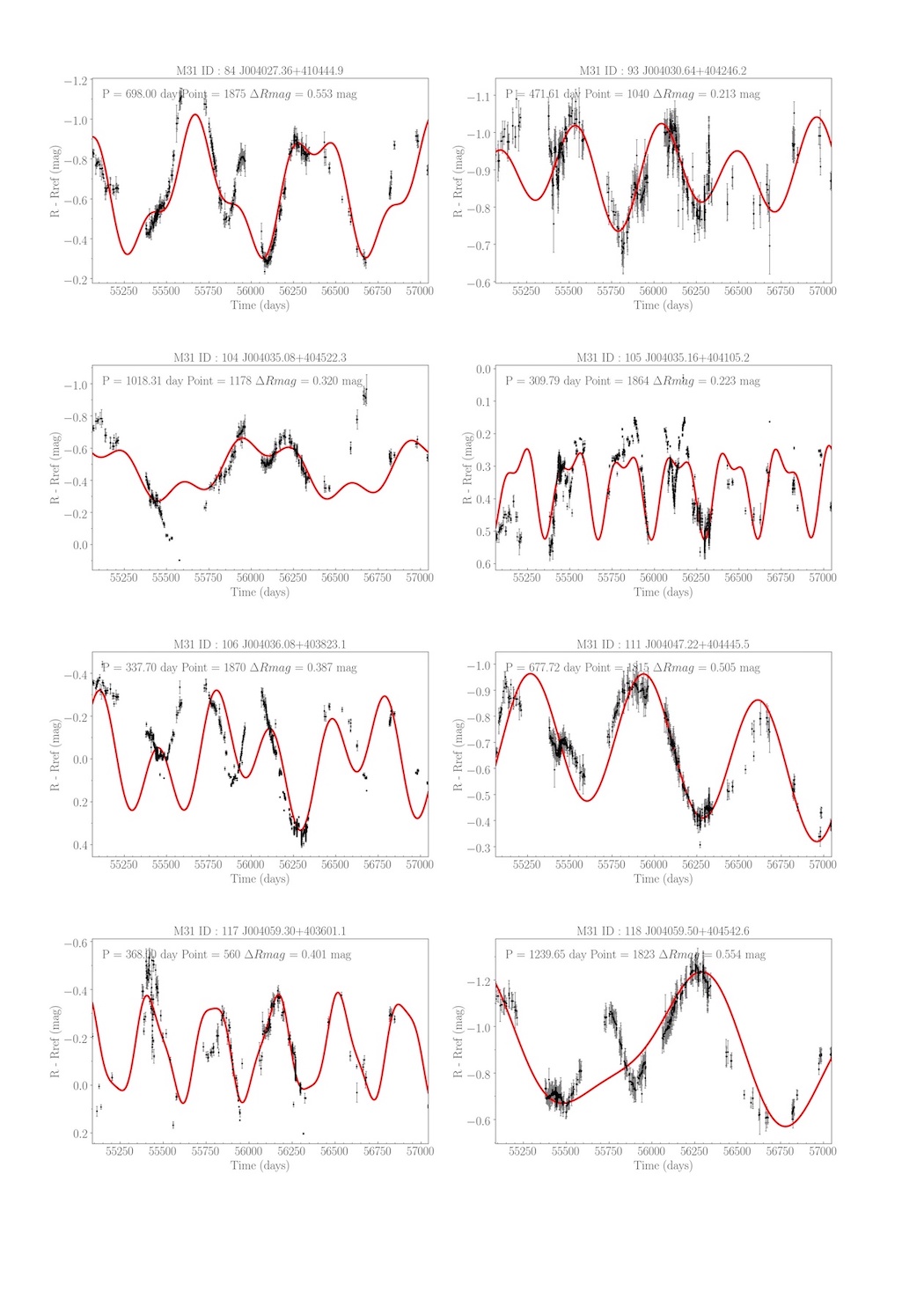}
\end{figure}

\begin{figure}[ht]
	\centering
	\includegraphics[width=\textwidth]{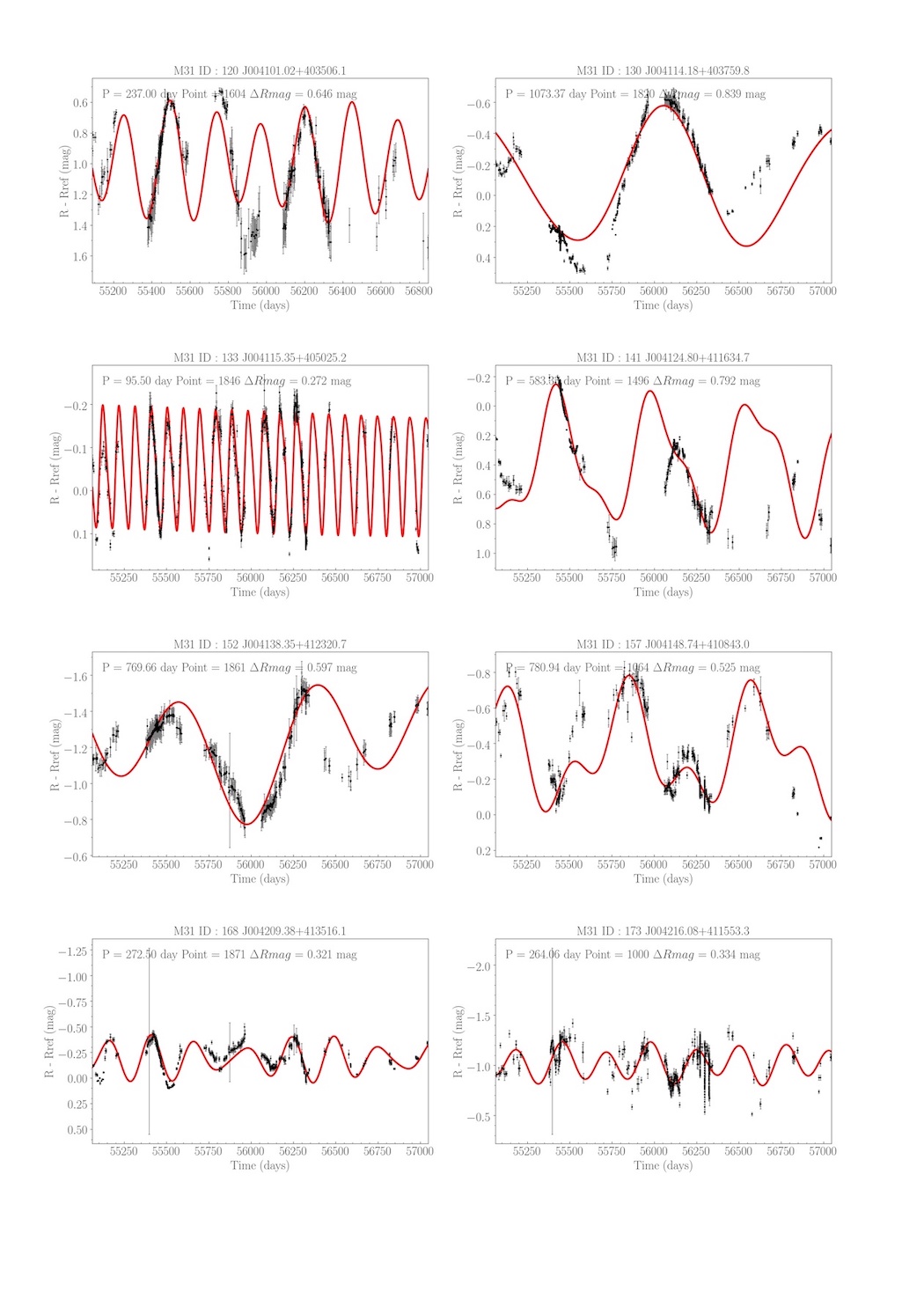}
\end{figure}

\begin{figure}[ht]
	\centering
	\includegraphics[width=\textwidth]{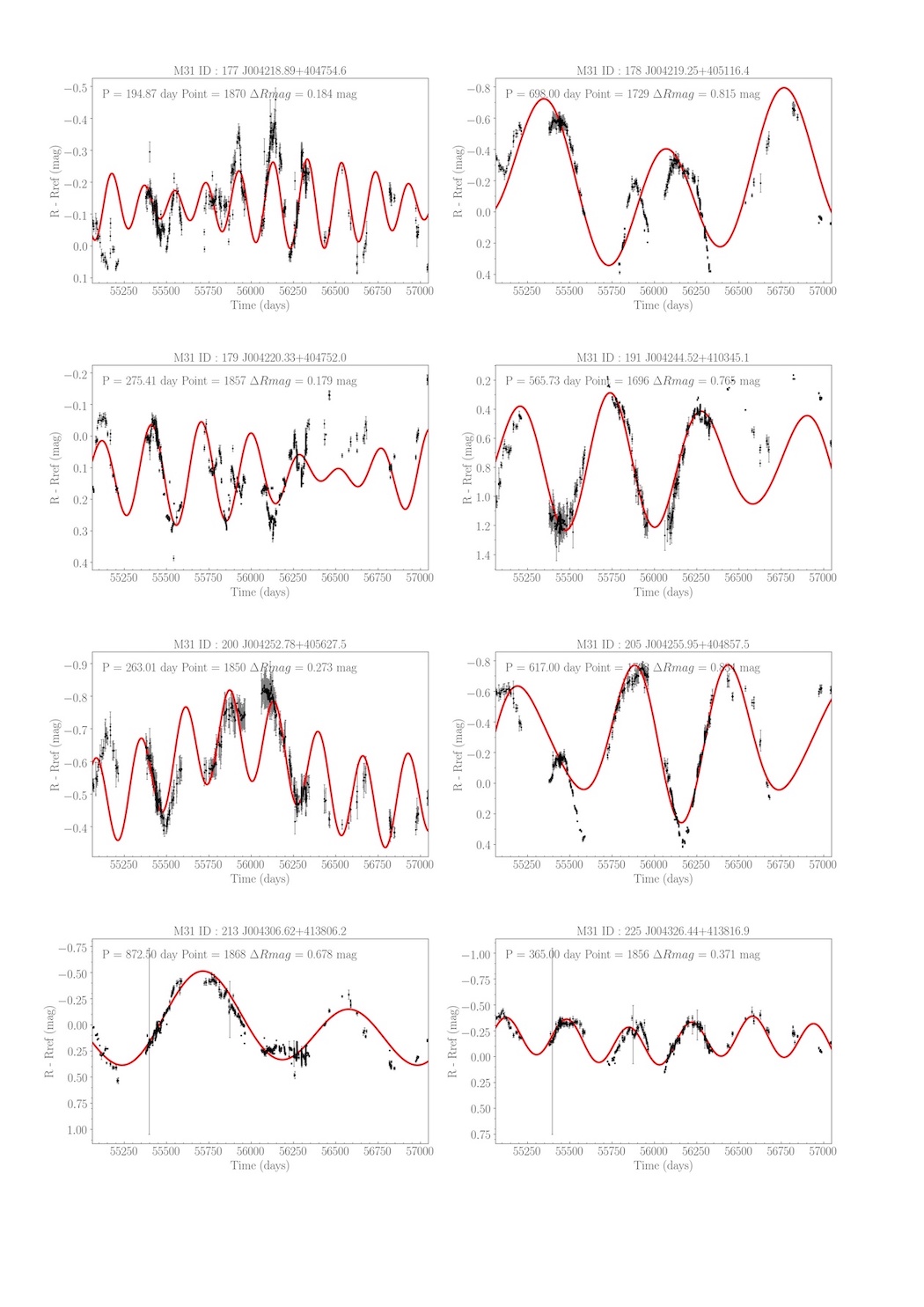}
\end{figure}

\begin{figure}[ht]
	\centering
	\includegraphics[width=\textwidth]{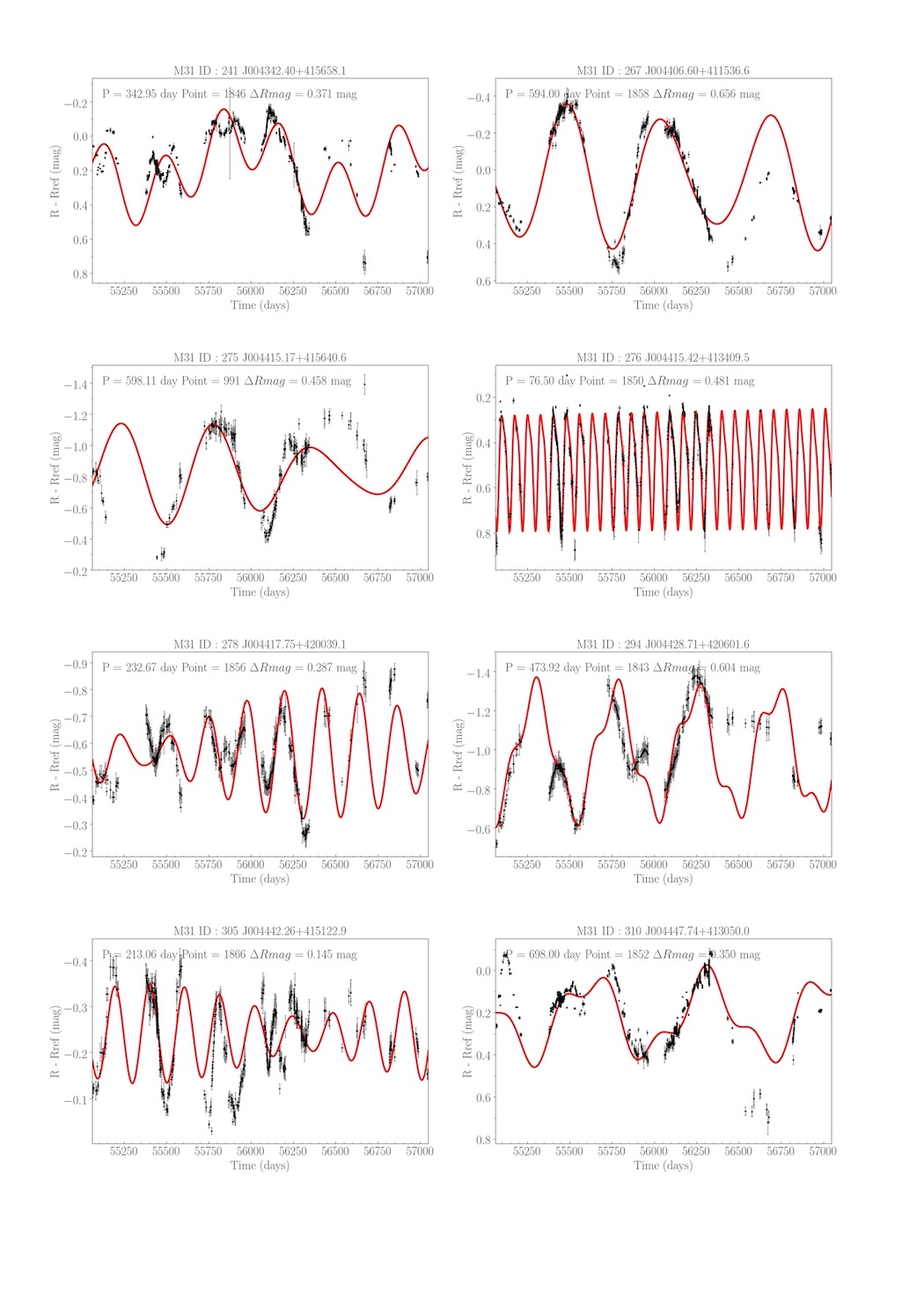}
\end{figure}

\begin{figure}[ht]
	\centering
	\includegraphics[width=\textwidth]{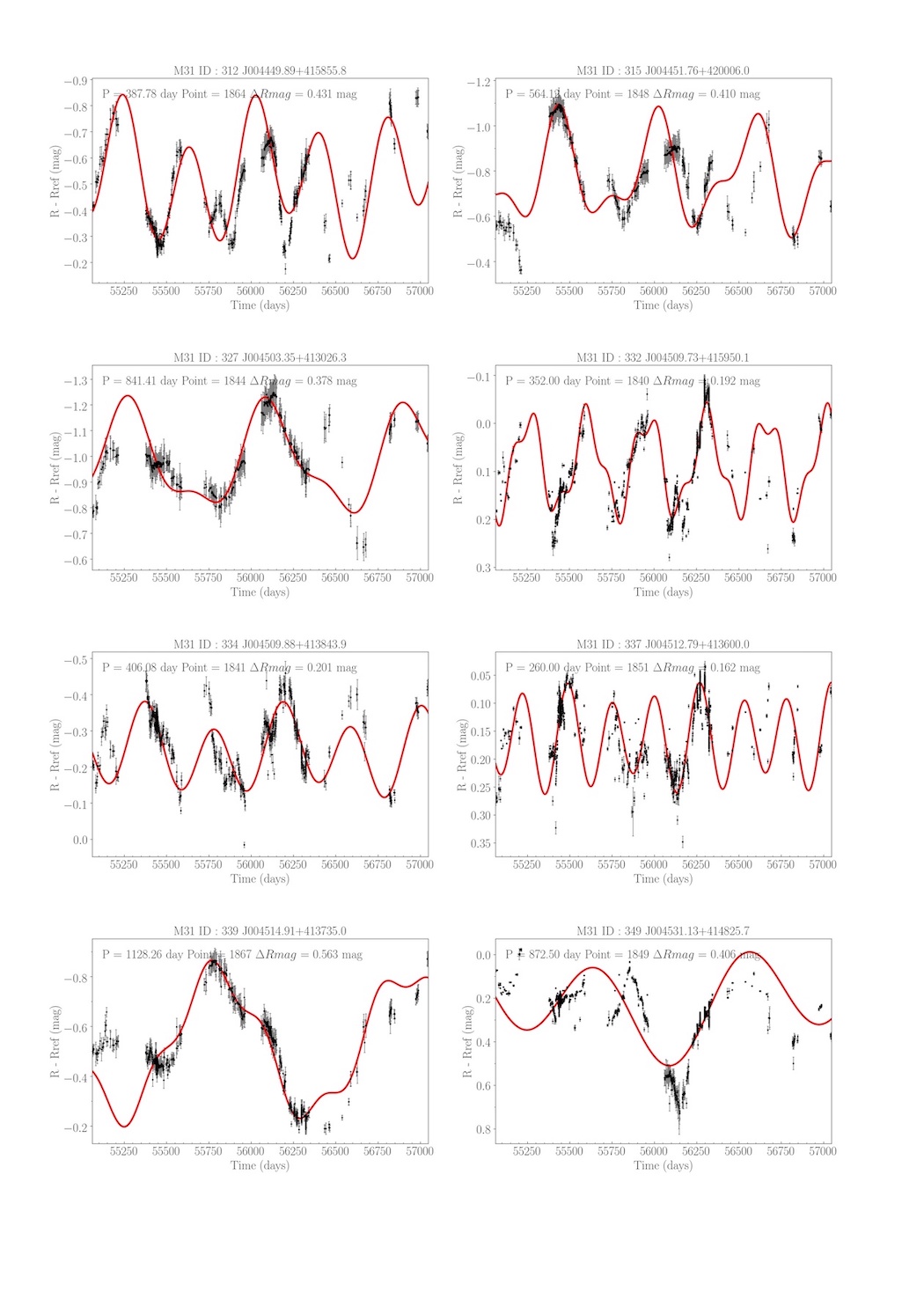}
\end{figure}

\begin{figure}[ht]
	\centering
	\includegraphics[width=\textwidth]{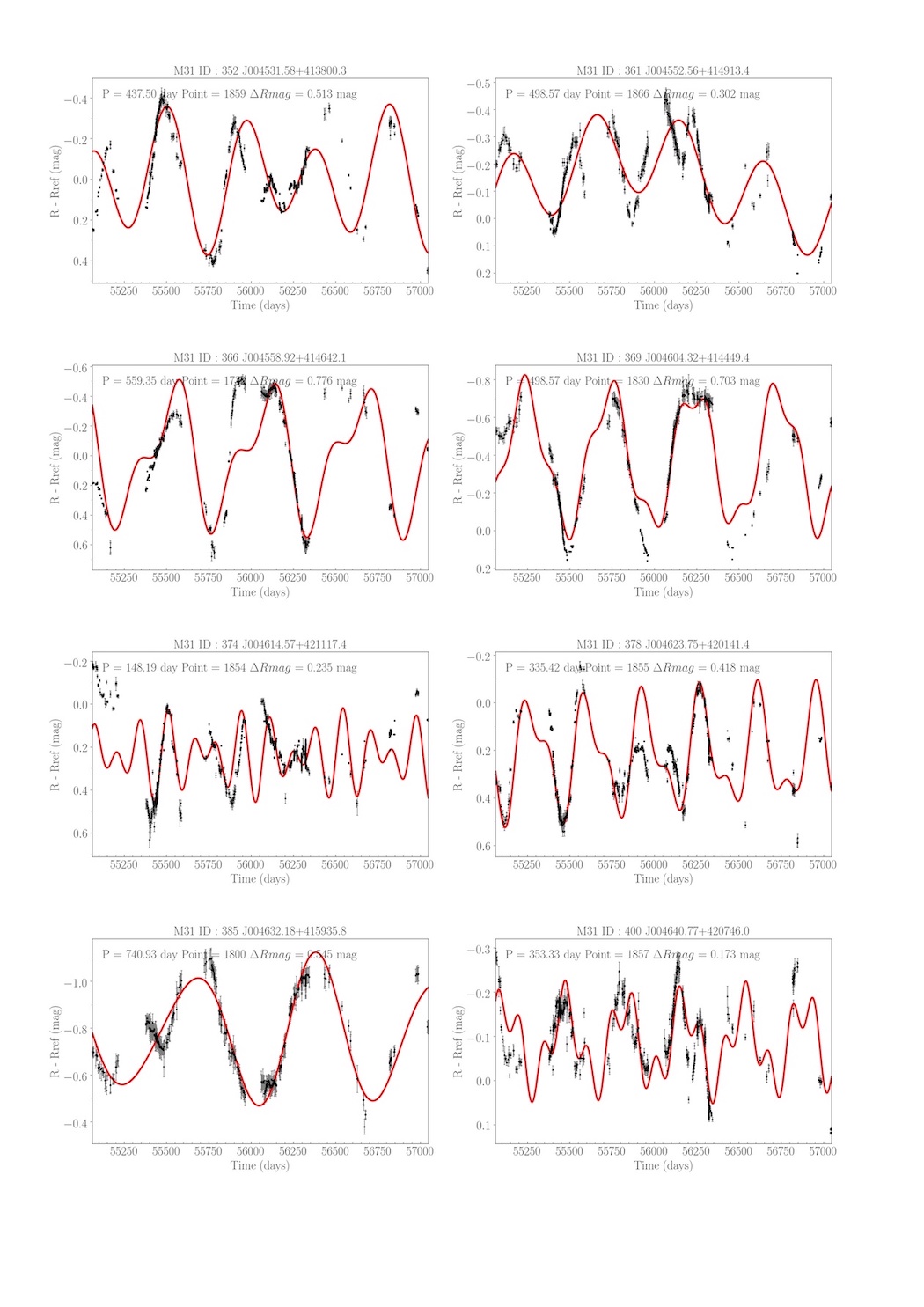}
\end{figure}

\section{Parameters of RSGs in M33 and M31} \label{appendix B}
\begin{longrotatetable}
\begin{deluxetable*}{ccccccccccccccccccc}
		
		\tablecaption{Parameters of RSGs in M33 \label{tab:Parameters of RSGs in M33}}
		\tablewidth{10pt}
		\tabletypesize{\scriptsize}
		\setlength{\tabcolsep}{1mm}
		
		\tablehead{
			\colhead{ID} & \colhead{LGGS} & \colhead{...} & \colhead{J} & \colhead{H} & \colhead{Ks} & \colhead{GLS} & \colhead{DFT} & \colhead{PDM} & \colhead{WWZ} & \colhead{Period} & \colhead{GLS-2} & \colhead{DFT-2} & \colhead{PDM-2} & \colhead{WWZ-2} & \colhead{Period-2} & \colhead{Amplitude} & \colhead{type} & \colhead{RMS}
		}
		\startdata
1 & J013147.31+302659.0 & ... & 16.72 & 15.982 & 15.786 & -- & -- & -- & -- & -- & -- & -- & -- & -- & -- & -- & -- & -- \\
2 & J013209.23+302614.8 & ... & 16.671 & 15.812 & 15.293 & -- & -- & -- & -- & -- & -- & -- & -- & -- & -- & -- & -- & -- \\
3 & J013213.60+304003.4 & ... & 16.76 & 16.162 & 15.1 & 386.808 & 77.799 & 187.5 & 91.842 & 77.799 & 33.802 & 232.955 & 187.5 & 317.273 & 317.273 & 0.117 & U & 0.1421 \\
4 & J013220.36+302938.0 & ... & -- & -- & -- & 881.409 & 793.248 & 1782.5 & 875 & 793.248 & 12.901 & 134.165 & 386 & 125 & 134.165 & 0.023 & N & 0.0715 \\
5 & J013222.59+302741.8 & ... & 16.669 & 15.746 & 15.77 & 34.058 & 15.841 & 2099 & 3500 & 15.841 & 3461.532 & 31.483 & 2099 & 1166.667 & 31.483 & 0.017 & N & 0.0743 \\
6 & J013223.56+303516.1 & ... & 16.706 & 15.863 & 15.113 & 71.546 & 444.812 & 1265 & 3500 & 3500 & 71.439 & 71.277 & 143 & 583.333 & 71.439 & 0.038 & N & 0.0727 \\
7 & J013223.63+302854.8 & ... & 16.746 & 15.832 & 16.098 & 443.786 & 455.454 & 1033 & 1166.667 & 455.454 & 15.622 & 14.17 & 1033 & 250 & 14.17 & 0.028 & N & 0.0725 \\
8 & J013224.33+303155.5 & ... & 16.742 & 16.114 & 15.572 & 14.446 & 121.03 & 305 & 318.182 & 121.03 & 118.619 & 312.61 & 305 & 700 & 312.61 & 0.022 & N & 0.0628 \\
\enddata
		\tablecomments{Table 6 is published in its entirety in the machine-readable format online.}
	\end{deluxetable*}
\end{longrotatetable}

\begin{longrotatetable}
	\begin{deluxetable*}{ccccccccccccccccccc}
		
		\tablecaption{Parameters of RSGs in M31 \label{tab:Parameters of RSGs in M31}}
		\tablewidth{10pt}
		\tabletypesize{\scriptsize}
		\setlength{\tabcolsep}{1mm}
		
		\tablehead{
			\colhead{ID} & \colhead{LGGS} & \colhead{...} & \colhead{J} & \colhead{H} & \colhead{Ks} & \colhead{GLS} & \colhead{DFT} & \colhead{PDM} & \colhead{WWZ} & \colhead{Period} & \colhead{GLS-2} & \colhead{DFT-2} & \colhead{PDM-2} & \colhead{WWZ-2} & \colhead{Period-2} & \colhead{Amplitude} & \colhead{type} & \colhead{RMS}
		}
		\startdata
1 & J003703.64+402014.9 & ... & 16.583 & 15.704 & 15.72 & -- & -- & -- & -- & -- & -- & -- & -- & -- & -- & -- & -- & -- \\
2 & J003721.98+401543.3 & ... & -- & -- & -- & -- & -- & -- & -- & -- & -- & -- & -- & -- & -- & -- & -- & -- \\
3 & J003722.34+400012.1 & ... & 16.045 & 15.368 & 15.026 & -- & -- & -- & -- & -- & -- & -- & -- & -- & -- & -- & -- & -- \\
4 & J003723.56+401715.5 & ... & 15.697 & 14.872 & 14.833 & -- & -- & -- & -- & -- & -- & -- & -- & -- & -- & -- & -- & -- \\
5 & J003724.48+401823.3 & ... & 16.236 & 15.006 & 15.11 & -- & -- & -- & -- & -- & -- & -- & -- & -- & -- & -- & -- & -- \\
6 & J003732.94+400015.8 & ... & 15.934 & 15.094 & 14.96 & -- & -- & -- & -- & -- & -- & -- & -- & -- & -- & -- & -- & -- \\
7 & J003739.41+395835.0 & ... & 16.146 & 15.359 & 15.132 & -- & -- & -- & -- & -- & -- & -- & -- & -- & -- & -- & -- & -- \\
8 & J003739.88+395838.7 & ... & 14.935 & 14.091 & 13.872 & -- & -- & -- & -- & -- & -- & -- & -- & -- & -- & -- & -- & -- \\
\enddata
		\tablecomments{Table 7 is published in its entirety in the machine-readable format online.}
	\end{deluxetable*}
\end{longrotatetable}




\begin{thebibliography}{}

\bibitem[Alcock et al.(1997)]{1997ApJ...486..697A} Alcock, C., Allsman, R. A., Alves, D., et al.,\ 1997, \apj, 486, 697
\bibitem[Astropy Collaboration et al.(2013)]{2013A&A...558A..33A} Astropy Collaboration, Robitaille, T.~P., Tollerud, E.~J., et al.\ 2013, \aap, 558, A33


\bibitem[Bertin \& Arnouts(1996)]{1996A&AS..117..393B} Bertin, E., \& Arnouts, S.\ 1996, \aaps, 117, 393
\bibitem[Bessell et al.(1998a)]{1998A&A...333..231B} Bessell, M. S., Castelli, F., \& Plez, B.,\ 1998, \aap, 333, 231
\bibitem[Bessell et al.(1998b)]{1998A&A...337..321B} Bessell, M. S., Castelli, F., \& Plez, B.,\ 1998, \aap, 337, 321
\bibitem[Bhardwaj et al.(2017)]{2017MNRAS.466.2805B} Bhardwaj, A. , Kanbur, S. M. , Marconi, M. , Rejkuba, M. , Singh, H. P. , \& Ngeow, C. C., \ 2017, MNRAS, 466, 2805
\bibitem[Brunish et al.(1986)]{1986AJ...91...598B} Brunish, W. M., Gallagher, J. S., \& Truran, J. W.,\ 1986, \aj, 91, 598


\bibitem[Cardelli et al.(1989)]{1989ApJ...345..245C} Cardelli, J. A., Clayton, G. C., \& Mathis, J. S.\ 1989, \apj, 345, 245
\bibitem[Chun et al.(2018)]{2018ApJ...853...79C} Chun, S., Yoon, S., Jung, M., Kim, D. U., \& Kim, J., \ 2018, \apj, 853, 79
\bibitem[Clayton et al.(2015)]{2015ApJ...815...14C} Clayton, G. C., Gordon, K. D., Bianchi, L. C., Massa, D. L., Fitzpatrick, E. L., \& Bohlin, R. C., et al., \ 2015, \apj, 815, 14

\bibitem[Drout et al.(2012)]{2012ApJ...750..97D} Drout, M. R., Massey, P., \& Meynet, G.,\ 2012, \apj, 750, 97
\bibitem[Dong et al.(2014)]{2014ApJ...785..136D} Dong, H., Li, Z., Wang, Q. D., et al., \ 2014, \apj, 785, 136
\bibitem[Dorda et al.(2016)]{2016A&A...592..16D} Dorda, R., Negueruela, I., Gonz Ã¡lez-Fern Ã¡ndez, C., \& Tabernero, H. M.,\ 2016, \aap, 592, 16

\bibitem[Ekstr\"{o}m et al.(2012)]{2012A&A...537..146E} Ekstr\"{o}m S., Georgy, C., Eggenberger, P., et al.,\ 2012, \aap, 537, 146
\bibitem[Elias et al.(1985)]{1985ApJS...57..91E} Elias, J. H., Frogel, J. A., \& Humphreys, R. M.,\ 1985, ApJS, 57, 91
\bibitem[Esteban \& Peimbert(1995)]{1995RMxAC...3..133E} Esteban, C., \& Peimbert, M.,\ 1995, MxAC, 3, 133

\bibitem[Gaia Collabortion et al.(2018)]{2018A&A...616A...1G} Gaia Collaboration, Brown, A. G. A., Vallenari, A., et al.,\ 2018, \aap, 616, 1
\bibitem[Garnett et al.(1997)]{1997ApJ...489..63G} Garnett, D. R., Shields, G. A., Skillman, E. D., Sagan, S. P., Dufour, R. J.,\ 1997, \apj, 489, 63
\bibitem[Guo \& Li(2002)]{2002ApJ...565..559G} Guo, J. H., \& Li, Y.\ 2002, \apj, 565, 559

\bibitem[Hartman et al.(2016)]{2016A&C....17....1H} Hartman, J. D., \& Bakos, G. A.,\ 2016, A\&C, 17, 1
\bibitem[Hog et al.(1998)]{1998A&A...335L..65H} Hog, E., Kuzmin, A., Bastian, U., Fabricius, C., Kuimov, K., Lindegren, L., et al.,\ 1998, \aap, 335, 65
\bibitem[Hughes \& Wood(1990)]{1990AJ.....99..784H} Hughes, S. M. G., \& Wood, P. R.,\ 1990, \aj, 99, 784

\bibitem[Josselin et al.(1996)]{1996AJ....112.1709F} Foster, G.\ 1996, \aj, 112, 1709
\bibitem[Josselin et al.(2000)]{2000A&A...357..225J} Josselin, E., Blommaert, J. A. D. L., Groenewegen, M. A. T., Omont, A., \& Li, F. L.\ 2000, \aap, 357, 225

\bibitem[Kinman et al.(1987)]{1987AJ.....93..833K} Kinman, T. D., Mould, J. R., Wood, P. R.,\ 1987, \aj, 93, 833
\bibitem[Kiss et al.(2006)]{2006MNRAS.372.1721K} Kiss, L. L., Szab\'{o}, Gy. M., Bedding, T. R., et al.\ 2006, MNRAS, 372, 1721
\bibitem[Kudritzki \& Reimers(1978)]{1978A&A....70..227K} Kudritzki, R-P., \& Reimers, D.\ 1978, \aap, 70, 227

\bibitem[Law et al.(2009)]{2009PASP..121.1395L} Law, N. M., Kulkarni, S. R., Dekany, R. G., et al.\ 2009, PASP, 121, 1395
\bibitem[Levesque et al.(2005)] {2005ApJ...628..973L} Levesque, E. M., Massey, P., Olsen, K. A. G., Plez, B., Josselin, E., Maeder, A., \& Meynet, G.\ 2005, \apj, 628, 973
\bibitem[Levesque et al.(2007)]{2007ApJ...667..202L} Levesque, E. M., Massey, P., Olsen, K. A. G., \& Plez, B. 2007, \apj, 667, 202
\bibitem[Levesque \& Massey(2012)]{2012AJ...144..2L} Levesque, E. M., \& Massey, P.,\ 2012, \aj, 144, 2
\bibitem[Lomb(1976)]{1976Ap&SS..39..447L} Lomb, N. R., \ 1976, ApSS, 39, 447

\bibitem[van Loon et al.(2005)]{2005A&A...438..273V} van Loon, J. Th., Cioni, M-R. L., Zijlstra, A. A., \& Loup, C.\ 2005, \aap, 438, 273

\bibitem[Maeder et al.(1980)]{1980A&A....90L..17M} Maeder, A., Lequeux, J., \& Azzopardi, M. 1980, \aap, 90, L17
\bibitem[Massey(1998)]{1998ApJ...501..153M} Massey, P.,\ 1998, \apj, 501, 153
\bibitem[Massey(2002)]{2002ApJS...141..81M} Massey, P.,\ 2002, ApJS, 141, 81
\bibitem[Massey \& Olsen(2003)]{2003AJ...126..2867M} Massey, P., \& Olsen, K. A. G.,\ 2003, \aj, 126, 2867
\bibitem[Massey et al.(2006)]{2006AJ....131.2478M} Massey, P., Olsen, K. A., Hodge, P. W., Strong, S. B., Jacoby, G. H., Schlingman, W. M., \& Smith, R. C.,\ 2006, \aj, 131, 2478
\bibitem[Massey et al.(2007)]{2007AJ....133..2393M} Massey, P., Olsen, K. A. G., Hodge, P. W., et al.,\ 2007, \aj, 133, 2393
\bibitem[Massey et al.(2008)]{2008IAUS..250...97M} Massey, P., Levesque, E. M., Plez, B., \& Olsen, K. A. G. 2008 in IAU Symp. 250, Massive Stars as Cosmic Engines, ed. F. Bresolin, P. Crowther, \& J. Puls (Cambridge: Cambridge Univ. Press), 97
\bibitem[Massey et al.(2009)]{2009ApJ...703..420M} Massey, P., Silva, D. R., Levesque, E. M., Plez, B., Olsen, K. A., Clayton, G. C., et al.\ 2009, \apj, 703, 420
\bibitem[Massey et al.(2016)]{2016AJ....152...62M} Massey, P., Neugent, K. F., \& Smart, B. M.\ 2016, \aj, 152, 62
\bibitem[Massey \& Evans(2016)]{2016ApJ...826..224M} Massey, P., \& Evans, K. A.\ 2016, \apj, 826, 224
\bibitem[Mauron \& Josselin(2011)]{2011A&A...526A.156M} Mauron, N., Josselin, E.\ 2011, \aap, 526, 156
\bibitem[Maund(2017)]{2017MNRAS.469.2202M} Maund, J. R.,\ 2017, MNRAS, 469, 2202

\bibitem[Orosz et al.(2007)]{2007Natur.449..872O} Orosz, J. A., Mcclintock, J. E., Narayan, R., Bailyn, C. D., Hartman, J. D., Macri, L. M., et al.,\ 2007, NATURE, 449, 872

\bibitem[Paxton et al.(2011)]{2011ApJS...192..3P} Paxton, B., Bildsten, L., Dotter, A., et al.,\ 2011, ApJS, 192, 3
\bibitem[Paxton et al.(2013)]{2013ApJS...208..4P} Paxton, B., Cantiello, M., Arras, P., et al.,\ 2013, ApJS, 208, 4
\bibitem[Paxton et al.(2015)]{2015ApJS...220..15P} Paxton, B., Marchant, P., Schwab, J., et al.,\ 2015, ApJS, 220, 15
\bibitem[Paxton et al.(2018)]{2018ApJS...234..34P} Paxton, B., Schwab, J., Bauer, E. B., et al.,\ 2018, ApJS, 234, 34
\bibitem[Perina et al.(2009)]{2009A&A...507.1375P} Perina, S., Federici, L., Bellazzini, M., Cacciari, C., Pecci, F. F., \& Galleti, S.\ 2009, \aap, 507, 1375
\bibitem[Pojmanski(2002)]{2002AcA....52..397P} Pojmanski, G.\ 2002, Acta Astron., 52, 397

\bibitem[Rau et al.(2009)]{2009PASP..121.1334R} Rau, A., Kulkarni, S. R., Law, N. M., et al.\ 2009, PASP, 121, 1334
\bibitem[Reimers(1975)]{1975MSRSL...8..369R} Reimers, D.\ 1975, MSRSL, 8, 369
\bibitem[Riechers et al.(2013)]{2013Natur.496..329R} Riechers, D. A., Bradford, C. M., Clements, D. L., Dowell, C. D., Perezfournon, I., Ivison, R. J., et al., \ 2013, Nature, 496, 329
\bibitem[Robin et al.(2003)]{2003A&A...409..523R} Robin, A. C., Reyl\'{e}, C., Derri\`{e}re, S., \& Picaud, S.\ 2003, \aap, 409, 523
\bibitem[Rubin \& Ford(1970)]{1970ApJ...159..379R} Rubin, V. C., \& Ford,W. K. J.\ 1970, \apj, 159, 379
\bibitem[Russell \& Dopita(1990)]{1990ApJS...74..93R} Russell, S. C., \& Dopita, M. C.,\ 1990 ,ApJS, 74, 93



\bibitem[Salvatier et al.(2016)]{2015arXiv150708050S} Salvatier, J., Wiecki, T. V., \& Fonnesbeck, C.2016, PeerJ Comp. Sci., 2, e55
\bibitem[Scargle(1982)]{1982ApJ...263..835S} Scargle, J. D.,\ 1982, ApJ, 263, 835
\bibitem[Smartt et al.(2009)]{2009MNRAS.395.1409S} Smartt, S. J., Eldridge, J. J., Crockett, R. M., \& Maund, J. R.,\ 2009, MNRAS, 395, 1409
\bibitem[Smartt(2015)]{2015PASA...32...16S} Smartt, S.,\ 2015, PASA, 32, 16
\bibitem[Smith \& Eichhorn(1996)]{1996MNRAS.281..211S} Smith H. Jr., Eichhorn H., Jul.\ 1996, MNRAS, 281, 211
\bibitem[Soraisam et al.(2018)]{2018ApJ...859...73S} Soraisam, M. D., Bildsten, L., Drout, M. R., Bauer, E. B., Gilfanov, M., \& Kupfer, T., et al., \ 2018, \apj, 859, 73
\bibitem[Stellingwerf(1978)]{1978ApJ...224..953S} Stellingwerf, R. F.\ 1978, \apj, 224, 953
\bibitem[Stothers \& Chin(1996)]{1996ApJ...468..842S} Stothers, R. B., \& Chin, C-W.\ 1996, \apj, 468, 842
\bibitem[Stothers \& Leung(1971)]{1971A&A....10..290S} Stothers, R., \& Leung, K. C.,\ 1971 ,\aap, 10, 290

\bibitem[Tammann et al.(2003)]{2003A&A...404..423T} Tammann, G. A., Sandage, A., \& Reindl, B., \ 2003, \aap, 404, 423

\bibitem[Verhoelst et al.(2009)]{2009A&A...498..127V} Verhoelst, T., Der Zypen, N. V., Hony, S., Decin, L., Cami, J., \& Eriksson, K.,\ 2009 ,\aap, 498, 127

\bibitem[Wang \& Jiang(2014)]{2014ApJL..788...12W} Wang, S, \& Jiang, B. W.,\ 2014, ApJL, 788, 12

\bibitem[Yang \& Jiang(2011)]{2011ApJ...727...53Y} Yang, M., \& Jiang, B. W.\ 2011, \apj, 727, 53
\bibitem[Yang \& Jiang(2012)]{2012ApJ...754...35Y} Yang, M., \& Jiang, B. W.\ 2012, \apj, 754, 35
\bibitem[Yang et al.(2018)]{2018A&A...616A.175Y} Yang, M., Bonanos, A. Z., Jiang, B. W., Gao, J., Xue, M., Wang, S., et al., \ 2018, \aap, 616, 175
\bibitem[Yecko et al.(1998)]{1998A&A...336..553Y} Yecko, P. A., Kollath, Z., \& Buchler, J. R., \ 1998, \aap, 336, 553

\bibitem[Zaritsky et al.(1994)]{1994ApJ...420..87Z}  Zaritsky, D., Kennicutt, R. C., Jr., \& Huchra, J. P.,\ 1994, \apj, 420, 87
\bibitem[Zechmeister \& Kurster(2009)]{2009A&A...496..577Z} Zechmeister, M., \& Kurster, M.,\ 2009, \aap, 496, 577


\end{thebibliography}
\end{document}